\documentclass[conference]{IEEEtran}

\usepackage[utf8]{inputenc}
\usepackage[T1]{fontenc}
\usepackage{amsmath,amssymb,amsfonts}
\usepackage{algorithmic}
\usepackage{algorithm}
\usepackage{graphicx}
\usepackage{textcomp}
\usepackage{xcolor}
\usepackage{booktabs}
\usepackage{multirow}
\usepackage{subcaption}
\usepackage{hyperref}
\usepackage{cleveref}
\usepackage{listings}
\usepackage{tikz}
\usepackage{pgfplots}
\usepackage{balance}

\pgfplotsset{compat=1.18}
\usetikzlibrary{arrows.meta, positioning, shapes.geometric, fit, calc, backgrounds}

\usepackage{amssymb}
\newcommand{\cmark}{\textcolor{green!60!black}{\checkmark}}

\definecolor{codegreen}{rgb}{0,0.6,0}
\definecolor{codegray}{rgb}{0.5,0.5,0.5}
\definecolor{codepurple}{rgb}{0.58,0,0.82}
\definecolor{backcolour}{rgb}{0.95,0.95,0.95}

\lstdefinestyle{verilog}{
    backgroundcolor=\color{backcolour},
    commentstyle=\color{codegreen},
    keywordstyle=\color{blue},
    numberstyle=\tiny\color{codegray},
    stringstyle=\color{codepurple},
    basicstyle=\ttfamily\scriptsize,
    breaklines=true,
    captionpos=b,
    keepspaces=true,
    numbers=left,
    numbersep=3pt,
    showspaces=false,
    showstringspaces=false,
    showtabs=false,
    tabsize=2,
    language=Verilog,
    morekeywords={module, endmodule, input, output, reg, wire, always, posedge,
                  negedge, assign, begin, end, if, else, case, endcase,
                  parameter, localparam, integer, genvar, generate, endgenerate}
}

\lstdefinestyle{python}{
    backgroundcolor=\color{backcolour},
    commentstyle=\color{codegreen},
    keywordstyle=\color{blue},
    numberstyle=\tiny\color{codegray},
    stringstyle=\color{codepurple},
    basicstyle=\ttfamily\scriptsize,
    breaklines=true,
    captionpos=b,
    keepspaces=true,
    numbers=left,
    numbersep=3pt,
    language=Python
}

\newcommand{\chipcraftbrain}{\textsc{ChipCraftBrain}}
\newcommand{\passatone}{pass@1}
\newcommand{\verilogeval}{\textsc{VerilogEval}}
\newcommand{\cvdp}{\textsc{CVDP}}
\newcommand{\mage}{\textsc{MAGE}}

\newcommand{\statedim}{168}

\newcommand{\mlparch}{Actor-critic MLP (shared 2$\times$256 trunk)}

\title{\chipcraftbrain{}: Validation-First RTL\\
Generation via Multi-Agent Orchestration}

\author{
\IEEEauthorblockN{Cagri Eryilmaz}
\IEEEauthorblockA{
ChipCraftX\\
San Francisco, CA\\
cagri@chipcraftx.io}
}

\begin{document}

\maketitle

\begin{abstract}
%

Large Language Models (LLMs) have shown promise for generating Register
Transfer Level (RTL) code from natural language specifications, yet
single-shot generation achieves only 60--65\% functional correctness on
standard benchmarks. Existing multi-agent approaches like \mage{} improve
pass rates to 95.9\% on \verilogeval{} but remain untested on harder
industrial benchmarks such as \cvdp{}, lack synthesis awareness, and
incur high API costs (\$0.06/problem with 20 parallel Sonnet calls).

We present \chipcraftbrain{}, a framework combining hybrid symbolic-neural
reasoning with adaptive multi-agent orchestration for automated RTL
generation. Four key innovations drive the system:
(1)~\textbf{Adaptive multi-agent orchestration} over six specialized
agents using a trained PPO policy over a 168-dimensional state
(with an alternative world-model MPC planner implemented and evaluated);
(2)~\textbf{Hybrid symbolic-neural architecture} that solves K-map and
truth table problems algorithmically (zero cost, perfect accuracy) while
using specialized agents for waveform timing analysis and general RTL
generation;
(3)~\textbf{Knowledge-augmented generation} via a curated domain knowledge
base of 321 patterns plus 971 open-source reference implementations
with focus-aware retrieval;
and (4)~\textbf{hierarchical specification decomposition} that breaks
complex multi-component designs into dependency-ordered sub-modules with
interface synchronization.

We evaluate across three complexity tiers.
On \verilogeval{}-Human (156 simple modules), \chipcraftbrain{} achieves
\textbf{98.7\%} \passatone{} (154/156) on its best run (range
96.15--98.72\% across 7 runs), placing it on par with or ahead of
ChipAgents (97.4\%, self-reported) and \mage{} (95.9\%) within
measurement noise.
On the 302-problem non-agentic, non-commercial code-generation subset
of NVIDIA's \cvdp{} benchmark, spanning five task categories (RTL
completion, spec-to-RTL, modification, linting, bug fixing), we
achieve \textbf{94.7\%} \passatone{} (286/302) under 5-iteration
refinement with error feedback. We lead all five categories against
the \cvdp{} paper's per-category single-shot baseline (Claude~3.7
Sonnet, 33.56\% aggregate), with per-category lifts of
36--60~percentage points; we additionally lead three of four
categories shared with NVIDIA's recent ACE-RTL agentic system despite
using roughly 30$\times$ fewer per-problem generation attempts.
A RISC-V SoC case study demonstrates hierarchical decomposition
generating 8/8 lint-passing modules (689~LOC), validated on FPGA
hardware, where monolithic generation fails entirely.

\end{abstract}



\section{Introduction}
\label{sec:introduction}

\subsection{The RTL Generation Challenge}

The semiconductor industry faces a persistent bottleneck: the demand for
custom chip designs far exceeds the supply of skilled RTL engineers.
As system-on-chip (SoC) complexity grows, with modern designs containing
billions of transistors across dozens of IP blocks, the time and expertise
required for front-end RTL design has become a critical constraint on
innovation.

Recent advances in Large Language Models (LLMs) have opened a promising
avenue: generating synthesizable Verilog or SystemVerilog code directly
from natural language specifications. However, the gap between
\emph{generating} code and \emph{generating correct, synthesizable} code
remains substantial. Single-shot generation with state-of-the-art models
achieves only 60--65\% pass rates on the \verilogeval{}
benchmark~\cite{liu2023verilogeval}, far below the reliability threshold
needed for production use.

Moreover, existing evaluations concentrate on simple benchmarks.
\verilogeval{} problems average 16 lines of code and 31 synthesis cells.
Real-world IP blocks are far more complex: NVIDIA's \cvdp{}
dataset~\cite{pinckney2025cvdp} spans 783 industrial problems across 13
task categories (code completion, spec-to-RTL, modification, linting,
bug fixing, testbench and assertion generation, and comprehension),
and ChipBench~\cite{yu2026chipbench} modules average 62 lines and 439
cells, 14$\times$ the gate count of \verilogeval{}. A system that excels
on simple modules but fails on industrial designs has limited practical
value.

\subsection{Limitations of Current Approaches}

Several lines of work have attempted to close this gap, each with
significant limitations:

\textbf{Single-shot generation} approaches~\cite{thakur2023verigen,
liu2024codev, liu2024rtlcoder} fine-tune LLMs on Verilog corpora but
provide no feedback mechanism. Errors in the generated code cannot be
detected or corrected, resulting in pass rates of 45--78\%.

\textbf{Iterative refinement} systems~\cite{thakur2023autochip} add
error-feedback loops but use fixed retry strategies that do not adapt
to the specific error patterns or design complexity of each problem.

\textbf{Multi-agent systems} like \mage{}~\cite{tsai2024mage} separate
RTL generation, testbench creation, judging, and debugging into
specialized agents. While achieving 95.9\% on \verilogeval{}-Human,
\mage{} has three critical limitations: (a)~no synthesis awareness in
candidate scoring (only functional correctness), (b)~primitive debug
feedback (raw waveform dumps), and (c)~high cost (20 parallel Claude
Sonnet calls per problem at \$0.06/problem). Crucially, while MAGE
reports 37.4\% on ChipBench, its performance on the broader \cvdp{}
benchmark has not been published.

\textbf{Workflow search} approaches like VFlow~\cite{vflow2025} use
Monte Carlo Tree Search to discover optimal agentic workflows, achieving
83.6\% pass@1. However, the discovered workflow is static; it does not
adapt per-problem based on design complexity or error patterns.

\subsection{Our Contributions}

We present \chipcraftbrain{}, a framework that addresses these limitations
through five key contributions:

\begin{enumerate}
    \item \textbf{Adaptive Multi-Agent Orchestration:} Six specialized
    LLM agents (four RL-orchestrated, two rule-based) coordinated by a
    PPO policy operating on a 168-dimensional state representation that
    captures specification complexity, error patterns, code metrics, and
    generation history (\Cref{sec:architecture}).

    \item \textbf{Hybrid Symbolic-Neural Architecture:} An algorithmic
    K-map solver using Quine-McCluskey minimization~\cite{quine1952problem,
    mccluskey1956minimization} handles deterministic problem classes at
    zero cost and perfect accuracy, while a specialized Waveform agent
    tackles temporal reasoning tasks (\Cref{sec:architecture}).

    \item \textbf{Knowledge-Augmented Generation:} A curated 321-entry
    domain knowledge base plus 971 open-source reference implementations with
    focus-aware RAG retrieval across five strategies, augmented by a
    spec-guidance registry of 59 detectors---akin to the heuristic
    design-pattern knowledge accumulated by commercial EDA tools---
    that enrich specifications before generation with
    problem-type-specific technical briefs
    (\Cref{sec:architecture}).

    \item \textbf{Hierarchical Specification Decomposition:} Automatic
    detection and decomposition of complex multi-component designs into
    4--8 dependency-ordered sub-modules with cross-module port
    synchronization, enabling generation of SoC-level designs where
    monolithic generation fails entirely (\Cref{sec:architecture}).

    \item \textbf{Three-Benchmark Evaluation Across Complexity Tiers:}
    We evaluate on \verilogeval{} (simple modules), \cvdp{} (industrial
    IP), and ChipBench (hard accelerator designs), demonstrating where
    current AI-driven RTL generation excels and where it breaks down
    (\Cref{sec:experiments}).
\end{enumerate}

\subsection{Paper Organization}

The remainder of this paper is organized as follows.
\Cref{sec:related} surveys related work.
\Cref{sec:architecture} presents the system architecture, including
multi-agent design, hybrid symbolic-neural components, RL orchestration,
knowledge retrieval, hierarchical decomposition, and visible reasoning.
\Cref{sec:validation} details the validation pipeline.
\Cref{sec:eval_framework} defines the three-benchmark evaluation framework.
\Cref{sec:experiments} reports experimental results and ablation studies.
\Cref{sec:case_study} presents a RISC-V SoC case study with FPGA
validation.
\Cref{sec:discussion} analyzes findings and limitations.
\Cref{sec:future} outlines future work, and
\Cref{sec:conclusion} concludes.


\section{Related Work}
\label{sec:related}

\subsection{LLM-Based RTL Generation}

The application of LLMs to hardware description language generation has
progressed rapidly. VeriGen~\cite{thakur2023verigen} demonstrated that
models fine-tuned on Verilog corpora can generate syntactically valid
code, though functional correctness remained limited.
CodeV~\cite{liu2024codev} improved on this with multi-level code
summaries for training, achieving 77.6\% on \verilogeval{}-Machine and
53.2\% on \verilogeval{}-Human using a 33B parameter model.
RTLCoder~\cite{liu2024rtlcoder} explored instruction-tuned models for
RTL, reaching 45--50\%. ChipChat~\cite{blocklove2023chipchat} applied
GPT-4 in conversational chip design workflows.
ChipNeMo~\cite{liu2023chipnemo} demonstrated that domain-adapted
pretraining on hardware-design corpora improves downstream task
performance but requires large labeled datasets.
More recently, OriGen~\cite{cui2024origen} enhances RTL generation
through code-to-code augmentation and self-reflection;
CraftRTL~\cite{liu2024craftrtl} targets correct-by-construction
synthetic data including non-textual design representations; and
ScaleRTL~\cite{deng2025scalertl} scales post-training on a 1.7\,M
RTL reasoning corpus, setting the current SoTA among standalone
RTL-specialized models.

These approaches share a fundamental limitation: without a feedback loop,
errors in the generated code cannot be detected or corrected, capping
practical accuracy.

\subsection{Multi-Agent Systems for RTL}

\mage{}~\cite{tsai2024mage} introduced a four-agent architecture
(RTL generator, testbench generator, judge, debugger) that achieves
95.9\% on \verilogeval{}-Human v2. Three innovations drive its success:
(a)~high-temperature sampling ($T=0.85$) with 20 candidates,
(b)~Verilog-state checkpoint debugging with textual waveform windows,
and (c)~multi-agent task separation that avoids context-switching
confusion between synthesizable RTL and non-synthesizable testbenches.
Their ablation shows the progression: vanilla LLM 72.4\% $\rightarrow$
single-agent 83.9\% $\rightarrow$ multi-agent 93.6\%.

However, \mage{}'s scoring function considers only functional correctness
(normalized mismatch count), ignoring synthesis quality metrics. Its debug
feedback is raw signal dumps without structural analysis. And generating
20 candidates with Claude Sonnet incurs substantial cost ($\sim$\$0.06
per problem). Critically, \mage{} has only been evaluated on
\verilogeval{}; its performance on harder industrial benchmarks is unknown.

ChipAgents~\cite{chipagents2025} reports 97.4\% on \verilogeval{}-v2
but has not published its methodology or evaluation protocol. As this
figure is self-reported without peer review, direct comparison should be
interpreted with caution.

\textbf{VerilogCoder}~\cite{ho2025verilogcoder} introduces a multi-agent
framework with a task-and-circuit relation graph for fine-grained task
decomposition and an AST-based waveform tracing tool for functional
repair. Its agent design inspired subsequent agentic systems but
depends on a custom Verilog parser that does not readily port to
SystemVerilog designs used in modern industrial benchmarks.

\textbf{ACE-RTL}~\cite{nvidia2026acertl} is NVIDIA's own agentic RTL
system, released concurrent with this work. It pairs a fine-tuned
Qwen2.5-Coder-32B generator (1.7\,M-sample RTL training corpus) with
Claude~4 Sonnet as reflector and coordinator, and runs 5 parallel
processes with up to 30 iterations each (up to 150 total attempts per
problem). It is the strongest published baseline to date on the
non-agentic \cvdp{} subset, and we report direct head-to-head
comparisons on the four categories shared with our evaluation in
\Cref{sec:cvdp_vs_acertl}.

\subsection{Fast Inference and Reasoning Models}

Beyond larger autoregressive models, two recent threads complement
pipeline-based RTL generation. Inception's \textsc{Mercury}
2~\cite{inception2025mercury} is a diffusion-based code LLM that
performs all-token refinement in parallel; routed through our pipeline
it achieves 82.7\% pass@1 on \verilogeval{}-Human at $\sim$5\,s/problem,
illustrating that non-autoregressive backbones can drive the same
iterative validation loop. Reasoning-optimized models such as
DeepSeek-R1~\cite{deepseek2025r1} and instruction-tuned coders such as
Qwen3-Coder~\cite{qwen2025qwen3coder} provide orthogonal gains on
long-horizon tasks. Our system treats the LLM as a replaceable
component, and the validation-first architecture benefits uniformly
from improvements at the backbone layer.

\subsection{Automated Workflow Discovery}

VFlow~\cite{vflow2025} applies Monte Carlo Tree Search (MCTS) to discover
optimal agentic workflows for Verilog generation, achieving 83.6\%
pass@1. The discovered five-step process (analysis $\rightarrow$
generation $\rightarrow$ ensemble $\rightarrow$ testing $\rightarrow$
refinement) incorporates Icarus Verilog syntax checking and Yosys
synthesis. However, the discovered workflow is static once found: it
does not adapt per-problem to design complexity or error patterns.

\subsection{RL for Code Generation}

Reinforcement learning has been applied to code generation in the
software domain. AlphaCode~\cite{li2022alphacode} uses large-scale
sampling and filtering for competitive programming.
CodeRL~\cite{le2022coderl} applies actor-critic methods for program
synthesis with execution feedback. However, no prior work applies RL
to orchestrate multi-agent \emph{hardware} code generation, where the
action space includes agent selection, temperature control, and
knowledge retrieval strategy, a fundamentally different problem due
to the synthesis and simulation constraints unique to RTL.

\subsection{RTL Generation Benchmarks}

\verilogeval{}~\cite{liu2023verilogeval} provides 156 problems
(average 16 LOC, 31 synthesis cells) and remains the most widely used
RTL benchmark, but saturation above 95\% limits its discriminative power.

NVIDIA's \cvdp{}~\cite{pinckney2025cvdp} is the most comprehensive
industrial RTL benchmark to date, with 783 problems authored by NVIDIA
hardware engineers across 13 task categories in two modes. The
non-agentic mode provides 617 single-turn problems spanning code
generation (cid002, cid003, cid004, cid007, cid016), testbench and
assertion generation (cid012, cid013, cid014), and comprehension
(cid006, cid008, cid009, cid010). The agentic mode provides 166
multi-file problems requiring Docker tool-use and iterative
refinement. The best published results on the 302-problem non-agentic
non-commercial code-generation subset are 33.56\% pass@1 with
Claude~3.7 Sonnet (single-shot, $n=5$ samples)~\cite{pinckney2025cvdp};
the agentic mode remains below 30\%. Testbench, assertion, and
commercial categories (cid012--cid014) require Cadence Xcelium and
thus cannot be replicated with open-source tooling.

ChipBench~\cite{yu2026chipbench} provides 45 Verilog generation problems
averaging 62 LOC and 439 cells, 3.8$\times$ longer and 14$\times$ more
gates than \verilogeval{}. It includes three categories of increasing
difficulty: self-contained modules (30), hierarchical designs requiring
sub-module instantiation (6), and RISC-V CPU IP components (9). On
ChipBench, the best multi-agent system (\mage{}) achieves only 37.4\%
and the best single model (Claude Opus) achieves 30.7\%, revealing the
gap between benchmark performance and real-world design capability.

\subsection{Hardware Design Automation}

Traditional EDA tools from Cadence (Cerebrus) and Synopsys (DSO.ai)
apply ML to \emph{backend} physical design optimization (place-and-route,
PPA)~\cite{synopsys_dso, cadence_cerebrus}. These are complementary to
front-end RTL generation. The OpenLane~2~\cite{openlane2} flow with
SKY130 PDK enables open-source RTL-to-GDSII synthesis, creating an
opportunity for end-to-end AI-driven chip design pipelines.


\section{System Architecture}
\label{sec:architecture}


\begin{figure*}[t]
\centering
\resizebox{\columnwidth}{!}{%
\begin{tikzpicture}[
    node distance=1.0cm and 1.2cm,
    >=stealth,
    input/.style={draw, rounded corners=3pt, fill=blue!8, minimum width=2.2cm,
                  minimum height=0.9cm, font=\small, align=center, thick},
    core/.style={draw, rounded corners=3pt, fill=orange!12, minimum width=2.4cm,
                 minimum height=1.0cm, font=\small, align=center, thick},
    agent/.style={draw, rounded corners=3pt, fill=green!10, minimum width=2.0cm,
                  minimum height=0.9cm, font=\small, align=center, thick},
    valid/.style={draw, rounded corners=3pt, fill=red!8, minimum width=2.0cm,
                  minimum height=0.9cm, font=\small, align=center, thick},
    output/.style={draw, rounded corners=3pt, fill=purple!8, minimum width=2.6cm,
                   minimum height=0.9cm, font=\small, align=center, thick},
    kb/.style={draw, rounded corners=3pt, fill=yellow!12, minimum width=2.2cm,
               minimum height=0.9cm, font=\small, align=center, thick},
    arrow/.style={->, thick, >=stealth},
    feedback/.style={->, thick, >=stealth, dashed, red!60!black},
    lbl/.style={font=\scriptsize, text=gray!70!black},
]

\node[input] (spec) {NL Spec\\{\scriptsize\texttt{name, description}}};

\node[core, below=0.8cm of spec] (router) {Complexity\\Router};

\node[core, below left=0.8cm and 2.5cm of router] (decomp) {Hierarchical\\Decomposition\\{\scriptsize 4--8 sub-modules}};
\node[core, below right=0.8cm and 2.5cm of router] (single) {Single-Pass\\Generation};

\node[core, below=2.4cm of router] (brain) {\textbf{ChipCraftBrain}\\{\scriptsize PPO Policy $\pi_\theta$}\\{\scriptsize \statedim{}-dim state $\to$ action}};

\node[kb, right=2.8cm of brain] (kb) {Knowledge\\Base\\{\scriptsize 321 entries}};

\node[agent, below=1.0cm of brain, xshift=-3.5cm] (fast) {Fast\\{\scriptsize Sonnet, $T$=0.5}};
\node[agent, below=1.0cm of brain, xshift=-1.2cm] (genius) {Genius\\{\scriptsize Opus, $T$=0.7}};
\node[agent, below=1.0cm of brain, xshift=1.2cm] (debug) {Debug\\{\scriptsize Opus, $T$=0.3}};
\node[agent, below=1.0cm of brain, xshift=3.5cm] (optimize) {Optimize\\{\scriptsize Sonnet, $T$=0.4}};

\coordinate[below=0.9cm of brain, yshift=-1.5cm] (merge);

\node[draw, rounded corners=3pt, fill=gray!8, minimum width=2.2cm,
      minimum height=0.7cm, font=\small, align=center, thick,
      below=2.5cm of brain] (code) {Verilog Code};

\node[valid, below=0.9cm of code, xshift=-3.5cm] (lint) {Lint\\{\scriptsize iverilog -t null}};
\node[valid, below=0.9cm of code] (sim) {Simulate\\{\scriptsize iverilog + vvp}};
\node[valid, below=0.9cm of code, xshift=3.5cm] (synth) {Synthesis\\{\scriptsize Yosys metrics}};

\node[output, below=0.9cm of sim] (out) {Verified RTL\\+ Reasoning Trace};

\draw[arrow] (spec) -- (router);
\draw[arrow] (router) -- node[lbl, left, pos=0.4] {$\geq$3 components} (decomp);
\draw[arrow] (router) -- node[lbl, right, pos=0.4] {simple} (single);
\draw[arrow] (decomp) |- (brain);
\draw[arrow] (single) |- (brain);

\draw[arrow] (brain) -- (fast);
\draw[arrow] (brain) -- (genius);
\draw[arrow] (brain) -- (debug);
\draw[arrow] (brain) -- (optimize);

\draw[arrow] (kb) -- node[lbl, above] {RAG} (brain);

\draw[arrow] (fast) -- (code);
\draw[arrow] (genius) -- (code);
\draw[arrow] (debug) -- (code);
\draw[arrow] (optimize) -- (code);

\draw[arrow] (code) -- (lint);
\draw[arrow] (code) -- (sim);
\draw[arrow] (code) -- (synth);

\draw[arrow, green!50!black] (lint) -- node[lbl, above] {pass} (sim);
\draw[arrow, green!50!black] (sim) -- node[lbl, above] {pass} (synth);

\draw[arrow] (synth) -- (out);

\draw[feedback] (lint.west) -- ++(-1.2, 0) |- node[lbl, left, pos=0.25]
    {errors + reward} (brain.west);

\begin{scope}[on background layer]
    \node[draw, dashed, gray!60, rounded corners=6pt, inner sep=10pt,
          fit=(brain)(fast)(optimize)(lint)(synth)(out)(code),
          label={[font=\scriptsize, gray!60]above left:RL-Guided Iterative Loop (max 5 iterations)}] {};
\end{scope}

\end{tikzpicture}%
}
\caption{%
    \chipcraftbrain{} system overview. Natural language specifications are
    routed through complexity analysis to either hierarchical decomposition
    or single-pass generation. The RL policy selects agents, temperatures,
    and RAG strategies. Generated Verilog passes through a three-stage
    validation pipeline; structured error feedback drives iterative refinement.%
}
\label{fig:system_overview}
\end{figure*}
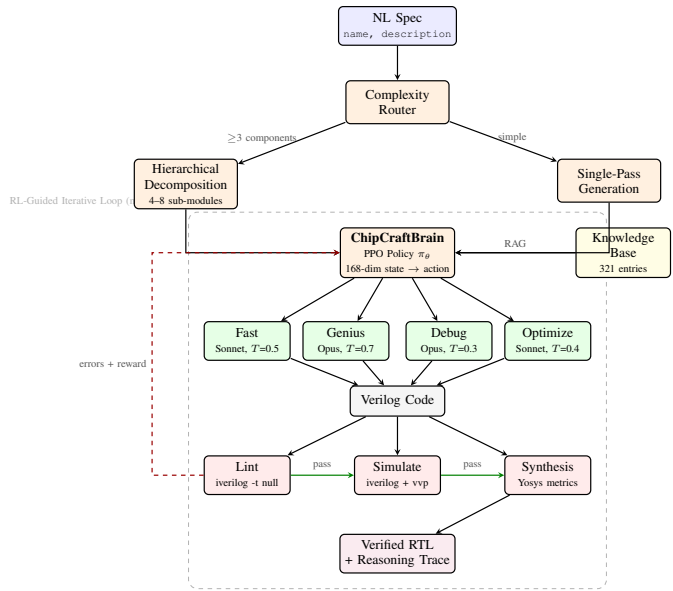

\subsection{System Overview}

\chipcraftbrain{} consists of six interconnected subsystems:
(1)~a \emph{multi-agent LLM system} with six specialized agents
(four RL-orchestrated, two rule-based),
(2)~a \emph{hybrid symbolic-neural system} providing algorithmic
solutions for K-map and truth table problems,
(3)~an \emph{RL orchestrator} that adaptively selects agents and
configures generation parameters,
(4)~a \emph{knowledge retrieval system} providing domain-specific
context via RAG,
(5)~a \emph{hierarchical decomposition engine} for complex designs, and
(6)~a \emph{validation pipeline} providing structured feedback.

The generation pipeline employs a three-tier selection strategy:
(1)~algorithmic solving for K-map and truth table problems (zero LLM cost),
(2)~rule-based routing for specialized problems (waveform timing analysis),
and (3)~RL-orchestrated selection among general agents for all other RTL
generation tasks. This hybrid architecture balances cost efficiency,
domain specialization, and learned adaptation.

\subsection{Multi-Agent LLM System}


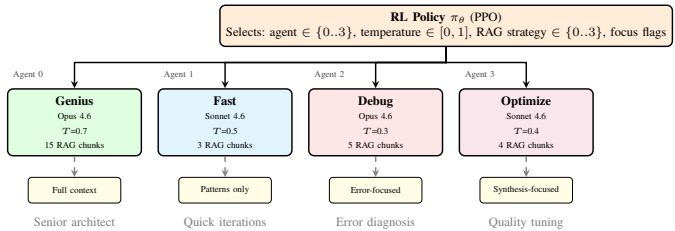
\begin{figure}[t]
\centering
\resizebox{\columnwidth}{!}{%
\begin{tikzpicture}[
    node distance=0.4cm and 0.4cm,
    >=stealth,
    agent/.style={draw, rounded corners=3pt, fill=#1, minimum width=2.6cm,
                  minimum height=1.3cm, font=\scriptsize, align=center, thick},
    controller/.style={draw, rounded corners=3pt, fill=orange!15,
                       minimum width=6.0cm, minimum height=0.8cm,
                       font=\scriptsize, align=center, thick},
    rag/.style={draw, rounded corners=2pt, fill=yellow!12, minimum width=1.8cm,
                minimum height=0.5cm, font=\tiny, align=center},
    arrow/.style={->, thick, >=stealth},
    label/.style={font=\tiny, text=gray!60!black},
]

\node[controller] (ctrl) {\textbf{RL Policy} $\pi_\theta$ (PPO)\\
    {\scriptsize Selects: agent $\in\{0..3\}$, temperature $\in [0,1]$,
     RAG strategy $\in\{0..3\}$, focus flags}};

\node[agent=green!12, below left=0.8cm and 1.5cm of ctrl] (genius) {%
    \textbf{Genius}\\
    {\tiny Opus 4.6}\\
    {\tiny $T$=0.7}\\
    {\tiny 15 RAG chunks}};

\node[agent=cyan!10, right=0.3cm of genius] (fast) {%
    \textbf{Fast}\\
    {\tiny Sonnet 4.6}\\
    {\tiny $T$=0.5}\\
    {\tiny 3 RAG chunks}};

\node[agent=red!10, right=0.3cm of fast] (debug) {%
    \textbf{Debug}\\
    {\tiny Opus 4.6}\\
    {\tiny $T$=0.3}\\
    {\tiny 5 RAG chunks}};

\node[agent=purple!10, right=0.3cm of debug] (optimize) {%
    \textbf{Optimize}\\
    {\tiny Sonnet 4.6}\\
    {\tiny $T$=0.4}\\
    {\tiny 4 RAG chunks}};

\node[rag, below=0.4cm of genius] (rag1) {Full context};
\node[rag, below=0.4cm of fast] (rag2) {Patterns only};
\node[rag, below=0.4cm of debug] (rag3) {Error-focused};
\node[rag, below=0.4cm of optimize] (rag4) {Synthesis-focused};

\draw[arrow] (ctrl.south) -- ++(0,-0.3) -| (genius.north);
\draw[arrow] (ctrl.south) -- ++(0,-0.3) -| (fast.north);
\draw[arrow] (ctrl.south) -- ++(0,-0.3) -| (debug.north);
\draw[arrow] (ctrl.south) -- ++(0,-0.3) -| (optimize.north);

\draw[arrow, dashed, gray] (genius) -- (rag1);
\draw[arrow, dashed, gray] (fast) -- (rag2);
\draw[arrow, dashed, gray] (debug) -- (rag3);
\draw[arrow, dashed, gray] (optimize) -- (rag4);

\node[label, above=0.05cm of genius.north west, anchor=south west] {Agent 0};
\node[label, above=0.05cm of fast.north west, anchor=south west] {Agent 1};
\node[label, above=0.05cm of debug.north west, anchor=south west] {Agent 2};
\node[label, above=0.05cm of optimize.north west, anchor=south west] {Agent 3};

\node[font=\scriptsize, text=gray, below=0.15cm of rag1]
    {Senior architect};
\node[font=\scriptsize, text=gray, below=0.15cm of rag2]
    {Quick iterations};
\node[font=\scriptsize, text=gray, below=0.15cm of rag3]
    {Error diagnosis};
\node[font=\scriptsize, text=gray, below=0.15cm of rag4]
    {Quality tuning};

\end{tikzpicture}%
}
\caption{%
    Multi-agent architecture. Four specialized LLM agents (plus
    Testbench and Waveform agents, not shown) serve distinct roles.
    The RL policy $\pi_\theta$ selects both the agent and its RAG
    retrieval strategy based on the \statedim{}-dimensional state vector.%
}
\label{fig:multi_agent}
\end{figure}

We employ six specialized agents, each with distinct prompts,
model configurations, and retrieval strategies:

\begin{table}[h]
\centering
\caption{Agent Specialization and Configuration}
\label{tab:agents}
\begin{tabular}{@{}llccl@{}}
\toprule
\textbf{Agent} & \textbf{Model} & \textbf{$T$} & \textbf{RAG $k$} & \textbf{Selection} \\
\midrule
\multicolumn{5}{l}{\textit{RL-Orchestrated (general-purpose):}} \\
Genius   & Opus 4.6   & 0.7 & 15 & RL policy \\
Fast     & Sonnet 4.6 & 0.5 & 3  & RL policy \\
Debug    & Opus 4.6   & 0.3 & 5  & RL policy \\
Optimize & Sonnet 4.6 & 0.4 & 4  & RL policy \\
\midrule
\multicolumn{5}{l}{\textit{Rule-Based (specialized):}} \\
Waveform  & Opus 4.6   & 0.4 & 15 & Keyword \\
Testbench & Opus 4.6   & 0.6 & 4  & Manual/API \\
\bottomrule
\end{tabular}
\end{table}

Following the key insight from \mage{}~\cite{tsai2024mage}, each agent
operates with an independent context window. Mixing RTL generation
(synthesizable code) and testbench creation (non-synthesizable code)
in the same conversation degrades both tasks. Our agents maintain
separate system prompts with role-specific instructions, including
Verilog coding guidelines, error patterns, and synthesis constraints.

The four general agents (Genius, Fast, Debug, Optimize) are selected
by the RL policy network based on problem characteristics and iteration
state. The Genius agent uses the most capable model (Opus) with deep
RAG context (15 chunks) for complex first attempts and novel designs.
The Fast agent (Sonnet) uses minimal context for quick refinement
iterations where latency matters. The Debug agent (Opus) specializes in
error-focused repair with targeted RAG retrieval on failure patterns.
The Optimize agent (Sonnet) focuses on synthesis quality improvement.
Role-specific models for Opus and Sonnet were chosen empirically: Opus
consistently outperforms Sonnet on error repair and initial generation,
while Sonnet offers 2--3$\times$ faster iterations at comparable
quality on well-specified refinement tasks.

The two specialized agents are activated via deterministic heuristics:
the Waveform agent is auto-selected when specifications contain temporal
reasoning keywords, and the Testbench agent is invoked explicitly for
test generation.

\subsection{Hybrid Symbolic-Neural Architecture}
\label{sec:hybrid}

Before invoking LLMs, we employ symbolic reasoning for problem classes
that admit deterministic solutions.

\subsubsection{Algorithmic K-map Solver}

Problems requesting Karnaugh map minimization or truth table implementation
are solved algorithmically using Quine-McCluskey
minimization~\cite{quine1952problem,mccluskey1956minimization}.
The solver:
\begin{itemize}
    \item Parses K-maps from natural language specifications (3--4 variables
    with Gray-coded columns)
    \item Extracts truth tables from ASCII tables or timing diagrams
    \item Computes minimal sum-of-products expressions
    \item Generates synthesizable Verilog-2001 code with appropriate
    bit-widths and signal declarations
    \item Handles don't-care conditions and detects XOR/XNOR patterns
\end{itemize}

Detection uses keyword matching (``karnaugh map'', ``k-map'', ``truth table'')
to route problems before LLM generation. On VerilogEval, this achieves
perfect accuracy (3/3 K-map problems) with \emph{zero API cost} and
\emph{zero iterations}.

\subsection{Specialized Waveform Analysis Agent}
\label{sec:waveform}

Timing waveform problems require precise temporal reasoning to infer
circuit behavior from signal transitions. These problems pose unique
challenges: (1)~subtle wrap-around conditions in counters (modulo-$N$
vs.\ bit-width overflow), (2)~ambiguous parallel vs.\ pipeline sampling
in dual-timing circuits (transparent latch + edge-triggered flip-flop),
and (3)~need for systematic trace analysis rather than pattern matching.

We introduce a dedicated \emph{Waveform agent} using Opus 4.6 with a
specialized system prompt that enforces a structured six-step reasoning
methodology:

\begin{enumerate}
    \item \textbf{Extract all transitions}: Enumerate (input, output)
    pairs at each time step from the waveform
    \item \textbf{Identify pattern type}: Classify as counter, state machine,
    or combinational logic
    \item \textbf{Find exact wrap points}: For counters, determine modulo
    value from observed sequence (not inferred from bit-width)
    \item \textbf{Resolve temporal dependencies}: For mixed latch+DFF circuits,
    identify signal flow (parallel sampling vs.\ pipeline chaining)
    \item \textbf{Derive logic}: Build if/else or case structure matching
    the observed behavior
    \item \textbf{Verify}: Check derived logic against multiple waveform samples
\end{enumerate}

This agent is auto-selected when specifications contain keywords
``waveform'', ``timing diagram'', or ``determine what the circuit does''.

\subsection{RL-Orchestrated Agent Selection}
\label{sec:rl}


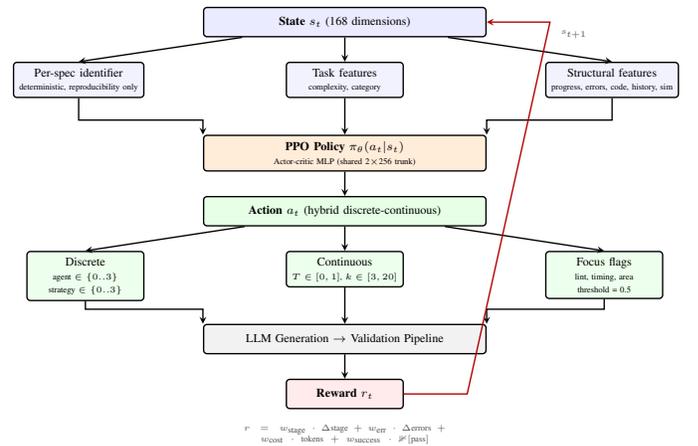
\begin{figure}[!b]
\centering
\resizebox{\columnwidth}{!}{%
\begin{tikzpicture}[
    node distance=0.5cm and 0.8cm,
    >=stealth,
    box/.style={draw, rounded corners=2pt, fill=#1, minimum width=2.4cm,
                minimum height=0.6cm, font=\scriptsize, align=center, thick},
    wide/.style={draw, rounded corners=2pt, fill=#1, minimum width=5.8cm,
                 minimum height=0.6cm, font=\scriptsize, align=center, thick},
    dim/.style={font=\tiny, text=gray!60!black},
    arrow/.style={->, thick, >=stealth},
    reward/.style={->, thick, >=stealth, red!70!black},
]

\node[wide=blue!8] (state) {\textbf{State} $s_t$ (\statedim{} dimensions)};

\node[box=blue!5, below left=0.5cm and 1.2cm of state] (s1)
    {Per-spec identifier\\{\tiny deterministic, reproducibility only}};
\node[box=blue!5, below=0.5cm of state] (s2)
    {Task features\\{\tiny complexity, category}};
\node[box=blue!5, below right=0.5cm and 1.2cm of state] (s3)
    {Structural features\\{\tiny progress, errors, code, history, sim}};

\node[wide=orange!15, below=2.0cm of state] (policy)
    {\textbf{PPO Policy} $\pi_\theta(a_t | s_t)$\\
     {\tiny \mlparch{}}};

\node[wide=green!10, below=0.5cm of policy]
    (action) {\textbf{Action} $a_t$ (hybrid discrete-continuous)};

\node[box=green!7, below left=0.5cm and 1.2cm of action] (a1)
    {Discrete\\{\tiny agent $\in\{0..3\}$}\\{\tiny strategy $\in\{0..3\}$}};
\node[box=green!7, below=0.5cm of action] (a2)
    {Continuous\\{\tiny $T \in [0,1]$, $k \in [3,20]$}};
\node[box=green!7, below right=0.5cm and 1.2cm of action] (a3)
    {Focus flags\\{\tiny lint, timing, area}\\{\tiny threshold = 0.5}};

\node[wide=gray!10, below=2.0cm of action] (env)
    {LLM Generation $\to$ Validation Pipeline};

\node[box=red!8, below=0.5cm of env] (rew)
    {\textbf{Reward} $r_t$};

\node[dim, below=0.2cm of rew, text width=5.5cm, align=center]
    {$r = w_{\text{stage}} \cdot \Delta\text{stage} + w_{\text{err}} \cdot
     \Delta\text{errors} + w_{\text{cost}} \cdot \text{tokens} +
     w_{\text{success}} \cdot \mathbb{1}[\text{pass}]$};

\draw[arrow] (state) -- (s1.north);
\draw[arrow] (state) -- (s2);
\draw[arrow] (state) -- (s3.north);

\draw[arrow] (s1.south) |- ([yshift=0.3cm]policy.north west) -- (policy.north west);
\draw[arrow] (s2) -- (policy);
\draw[arrow] (s3.south) |- ([yshift=0.3cm]policy.north east) -- (policy.north east);

\draw[arrow] (policy) -- (action);

\draw[arrow] (action) -- (a1.north);
\draw[arrow] (action) -- (a2);
\draw[arrow] (action) -- (a3.north);

\draw[arrow] (a1.south) |- ([yshift=0.3cm]env.north west) -- (env.north west);
\draw[arrow] (a2) -- (env);
\draw[arrow] (a3.south) |- ([yshift=0.3cm]env.north east) -- (env.north east);

\draw[arrow] (env) -- (rew);

\coordinate (fb1) at ([xshift=1.3cm]rew.east);
\coordinate (fb2) at ([xshift=1.3cm]state.east);
\draw[reward] (rew.east) -- (fb1) -- (fb2) -- (state.east);
\node[dim, anchor=west] at ([xshift=0.1cm,yshift=-0.25cm]fb2) {$s_{t+1}$};

\end{tikzpicture}%
}
\caption{%
    RL orchestration loop. The \statedim{}-dimensional state vector
    combines structural features (task metadata, generation
    progress, errors, code metrics, agent history, simulator
    signals) with a deterministic per-spec identifier used only for
    reproducibility. The PPO policy outputs a hybrid action
    selecting agent, RAG strategy, temperature, and focus flags. A
    multi-component reward drives learning from validation
    outcomes.%
}
\label{fig:rl_loop}
\end{figure}

For general RTL generation tasks (after algorithmic and rule-based routing),
we employ reinforcement learning to adaptively select among four specialized
agents and configure generation parameters.

\subsubsection{State Representation}

We encode the generation state as a 168-dimensional vector
combining forty structural features --- task metadata
(complexity, category); generation progress (iteration state and
structural-refinement phase context); structured error signals;
code-level metrics; agent-selection history; and simulator
performance observations --- with a deterministic per-spec
identifier computed once from the specification text. The
identifier provides reproducible policy behavior across machines
and training runs; generalization across unseen specifications
flows through the structural features.

\subsubsection{Action Space}

The policy selects a hybrid discrete-continuous action:
\begin{equation}
\mathbf{a} = (\underbrace{a_{\text{agent}}, a_{\text{focus}}}_{\text{discrete}},
\underbrace{a_{T}, a_{\text{tokens}}, a_{\text{rag}}, a_{\text{retry}}}_{\text{continuous}})
\label{eq:action}
\end{equation}

\noindent where $a_{\text{agent}} \in \{$Genius, Fast, Debug, Optimize$\}$,
$a_{\text{focus}} \in \{$full, minimal, error, synthesis, architecture$\}$,
$a_T \in [0, 1]$ (sampling temperature),
$a_{\text{tokens}} \in [0, 1]$ (normalized token budget),
$a_{\text{rag}} \in [0, 1]$ (RAG depth),
and $a_{\text{retry}} \in [0, 1]$ (retry budget allocation).

\subsubsection{Reward Function}

The multi-component reward balances correctness, efficiency, and quality:
\begin{equation}
R = R_{\text{term}} + R_{\text{eff}} + R_{\text{qual}} + R_{\text{prog}}
\label{eq:reward}
\end{equation}

\noindent where:
$R_{\text{term}} = \{+100\text{ sim pass}, +60\text{ lint pass},
-50\text{ total failure}\}$,
$R_{\text{eff}} = \{+20\text{ first try}, -0.001/\text{token}\}$,
$R_{\text{qual}} = \{+10\text{ low LUT}, +15\text{ timing met}\}$,
and $R_{\text{prog}} = \{+5/\text{stage advance},
+3/\text{error eliminated}\}$.

\subsubsection{Policy Network}
\label{eq:policy}

The policy is an actor-critic network trained via Proximal Policy
Optimization (PPO)~\cite{schulman2017ppo}. A shared two-layer MLP
trunk (each layer 256-wide with LayerNorm and ReLU) maps the
168-dimensional state to a 256-dim feature; three linear heads then
emit the discrete-action logits, continuous-action parameters
(sigmoid-bounded), and the scalar value estimate. Training uses
GAE ($\lambda=0.95$, $\gamma=0.99$), the clipped surrogate
objective, and epsilon-greedy exploration ($\varepsilon_0=0.3$,
decay $0.995$/episode). The first 20 episodes use heuristic
fallback for warm-start.

\subsubsection{Training Regime}
\label{sec:rl_training_regime}

The policy network is initialized through a two-phase procedure.
During the first twenty episodes, action selection is governed by a
rule-based heuristic that maps error type and iteration stage to
agent, temperature, and retrieval configuration; these episodes
populate an experience buffer and serve as a warm start. From
episode twenty onward, online PPO updates are applied using the
clipped surrogate objective with experience replay. The released
checkpoint corresponds to a policy trained through fifty-seven
episodes with the exploration rate decayed from $\varepsilon_0=0.3$
to $\varepsilon=0.225$; the network architecture is as described
in the Policy Network subsection above.

Because the online training set overlaps with \verilogeval{}-Human,
which also serves as an evaluation benchmark, pass-rate gains that
are attributable to the trained policy cannot be cleanly separated
from potential memorization of individual problems. We therefore
regard the trained policy as a proof of concept rather than a
generalization claim, and defer clean train--test separation to a
held-out corpus of pattern-mined and synthetically-generated
specifications disjoint from \verilogeval{}, \cvdp{}, and ChipBench.
This separation, together with a leave-one-out ablation against the
heuristic warm-start policy, is reported as future work
(\Cref{sec:future}).

As an alternative to the learned PPO policy, the system supports a
model-predictive-control (MPC) planner that rolls out sixty-four
candidate action sequences over a three-step horizon against a
trained RTL world model. On a controlled comparison over the 156
\verilogeval{}-Human problems, the MPC planner attains a pass rate
of $97.4\%$ versus $96.8\%$ for the trained PPO policy under matched
inference budgets, with a $1.8\%$ reduction in average iterations per
problem and no observed regressions. The availability of two
planners with similar end-to-end behavior on this benchmark supports
the interpretation that the iterative validation loop and agent
specialization contribute more to the reported accuracy than the
specific planner used to route between agents. The headline result
of $98.7\%$ reported in \Cref{tab:verilogeval_results} corresponds to
the best of seven full-pipeline runs (range $96.15$--$98.72\%$); the
planner comparison above is drawn from a single controlled run, and
cross-run variance is discussed in \Cref{sec:experiments}.

\subsection{Knowledge-Augmented Generation}
\label{sec:rag}


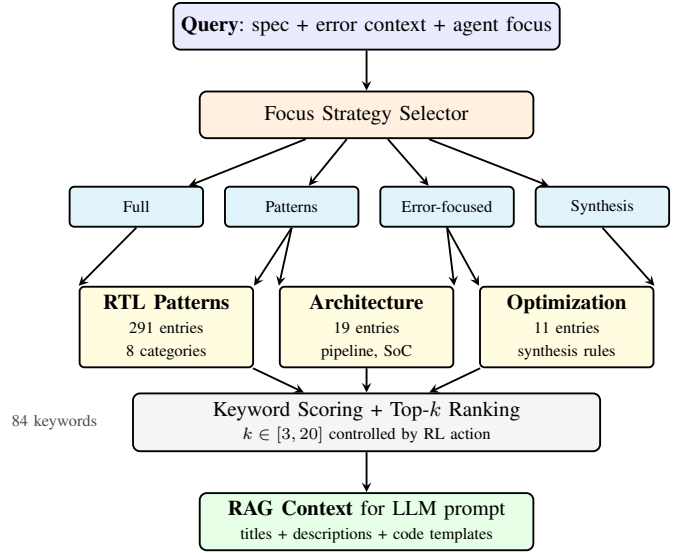
\begin{figure}[t]
\centering
\resizebox{\columnwidth}{!}{%
\begin{tikzpicture}[
    node distance=0.5cm and 0.6cm,
    >=stealth,
    kb/.style={draw, rounded corners=2pt, fill=yellow!15, minimum width=2.6cm,
               minimum height=1.2cm, font=\small, align=center, thick},
    focus/.style={draw, rounded corners=2pt, fill=cyan!10, minimum width=2.0cm,
                  minimum height=0.6cm, font=\scriptsize, align=center, thick},
    query/.style={draw, rounded corners=3pt, fill=blue!8, minimum width=5.0cm,
                  minimum height=0.7cm, font=\small, align=center, thick},
    output/.style={draw, rounded corners=3pt, fill=green!10, minimum width=5.0cm,
                   minimum height=0.7cm, font=\small, align=center, thick},
    arrow/.style={->, thick, >=stealth},
    label/.style={font=\scriptsize, text=gray!60!black},
]

\node[query] (q) {\textbf{Query}: spec + error context + agent focus};

\node[draw, rounded corners=3pt, fill=orange!12, minimum width=5.0cm,
      minimum height=0.7cm, font=\small, align=center, thick,
      below=0.6cm of q] (selector) {Focus Strategy Selector};

\node[focus, below=0.7cm of selector, xshift=-3.45cm] (f1) {Full};
\node[focus, right=0.3cm of f1] (f2) {Patterns};
\node[focus, right=0.3cm of f2] (f3) {Error-focused};
\node[focus, right=0.3cm of f3] (f4) {Synthesis};

\node[kb, below=2.2cm of selector, xshift=-3.0cm] (pat) {%
    \textbf{RTL Patterns}\\
    {\scriptsize 291 entries}\\
    {\scriptsize 8 categories}};
\node[kb, below=2.2cm of selector] (arch) {%
    \textbf{Architecture}\\
    {\scriptsize 19 entries}\\
    {\scriptsize pipeline, SoC}};
\node[kb, below=2.2cm of selector, xshift=3.0cm] (opt) {%
    \textbf{Optimization}\\
    {\scriptsize 11 entries}\\
    {\scriptsize synthesis rules}};

\node[draw, rounded corners=3pt, fill=gray!8, minimum width=7.0cm,
      minimum height=0.7cm, font=\small, align=center, thick,
      below=3.8cm of selector] (rank) {Keyword Scoring + Top-$k$ Ranking\\
      {\scriptsize $k \in [3, 20]$ controlled by RL action}};

\node[output, below=0.6cm of rank] (out) {\textbf{RAG Context} for LLM prompt\\
      {\scriptsize titles + descriptions + code templates}};

\draw[arrow] (q) -- (selector);
\draw[arrow] (selector) -- (f1);
\draw[arrow] (selector) -- (f2);
\draw[arrow] (selector) -- (f3);
\draw[arrow] (selector) -- (f4);

\draw[arrow] (f1.south) -- (pat.north west);
\draw[arrow] (f2.south) -- (pat.north east);
\draw[arrow] (f2.south) -- (arch.north west);
\draw[arrow] (f3.south) -- (arch.north east);
\draw[arrow] (f3.south) -- (opt.north west);
\draw[arrow] (f4.south) -- (opt.north east);

\draw[arrow] (pat) -- (rank);
\draw[arrow] (arch) -- (rank);
\draw[arrow] (opt) -- (rank);

\draw[arrow] (rank) -- (out);

\node[label, left=0.4cm of rank] {84 keywords};

\end{tikzpicture}%
}
\caption{%
    Knowledge-augmented retrieval. The 321-entry knowledge base spans
    three sources (RTL patterns, architecture, optimization). The RL
    policy selects a focus strategy that weights sources differently;
    keyword-based scoring ranks candidates, and the top-$k$ chunks
    (where $k$ is an RL-controlled parameter) are injected into the
    LLM prompt.%
}
\label{fig:knowledge_rag}
\end{figure}

\chipcraftbrain{} maintains a multi-source knowledge system:

\textbf{Curated Knowledge Base.}
321 entries spanning three categories: \textbf{RTL design patterns}
(291 entries covering combinational logic, sequential circuits, FSMs,
memory controllers, bus protocols, CDC, and DSP),
\textbf{architecture templates} (19 entries for CPU cores, caches,
interconnects, SoC integration), and \textbf{optimization strategies}
(11 entries for area, timing, power, and FPGA techniques).

\textbf{Reference Implementation Library.}
Complementing the curated knowledge base, we assemble a secondary
corpus of 971 indexed reference modules harvested from 34
open-source Verilog and SystemVerilog repositories, all
permissively licensed and redistributed with attribution. Each
module is parsed, syntax-checked with Icarus~Verilog, and registered
in a per-category index with its port list, parameterization, and a
short natural-language synopsis; the indexer discards files that
fail lint or exceed a size threshold before retrieval. The library
is partitioned into fourteen hardware domains, listed in
\Cref{tab:reference_library}. Upstream sources include the Adapteva
\textsc{OpenHardware} collection, Alex Forencich's high-speed
interface projects, the OpenRISC and RISC-V open-core ecosystems,
and several widely-used cryptographic, arithmetic, and interconnect
cores; no single source contributes more than one-sixth of the
library. At retrieval time, the library is ranked by
the same focus-aware strategies as the curated knowledge base and
may be drawn from in place of, or alongside, the curated entries.

\begin{table}[h]
\centering
\caption{Reference implementation library: entries per hardware
domain. All modules are open source and were filtered for
synthesizability before indexing.}
\label{tab:reference_library}
\begin{tabular}{@{}lrlr@{}}
\toprule
\textbf{Domain} & \textbf{N} & \textbf{Domain} & \textbf{N} \\
\midrule
CPU building blocks   & 174 & Memory                & 37  \\
Datapath              & 165 & Cryptographic cores   & 33  \\
Protocol bridges      & 153 & System infrastructure & 33  \\
Network / packet      & 98  & Clock and reset       & 20  \\
Peripheral controllers& 78  & SerDes / line coding  & 18  \\
DSP                   & 77  & Arbitration           & 11  \\
Flow control          & 65  & Error correction      & 9   \\
\midrule
\multicolumn{3}{l}{\textbf{Total}} & \textbf{971} \\
\bottomrule
\end{tabular}
\end{table}

\textbf{Spec Guidance Registry.}
The registry enriches a specification with problem-type-specific
technical guidance before generation. The entry point,
\texttt{enrich\_spec()}, is invoked once at the top of
\texttt{generate\_module()}, so every code path (CLI, benchmark
runner, programmatic API, local-model backend) receives identical
enrichment. Input is a raw natural-language spec; output is the same
spec with one or more guidance blocks appended.

The registry contains 59 detectors organized into five thematic
bands: interface preservation for RTL modification tasks; classic
RTL hazards (timing, latency, edge capture, waveform-to-logic);
structural protocol patterns (APB, AXI-Stream, skid buffers, TLBs,
interrupt controllers); microarchitecture idioms (NoC routers,
systolic arrays, DRAM controllers, CPU-IP components); and
machine-learning operator primitives. Detectors are evaluated
in priority order, and enrichment is cumulative: every match
appends its guidance block. Priority-sensitive dispatch is a
first-class concern; for example, the APB-specific interrupt
controller detector precedes the generic priority detector, and the
attention detector precedes matmul and reduce, because attention
specifications mention matrix multiplication but require additional
invariants.

Two detector classes coexist. \emph{Semantic} detectors trigger on
pattern conjunctions in the description (e.g., ``start bit'' and
``stop bit'' implies a serial FSM; ``neighbour'' or ``neighbor''
with a width indicator implies bit-neighbor combinational logic).
\emph{Fixture-grounded} detectors trigger on named IP families that
the base model reliably fails on without structural scaffolding
(e.g., APB interrupt controllers, Sony IR decoders, specific
memory-allocator interfaces); targeted guidance for these families
is more reliable than open-ended semantic heuristics. Each guidance
block is a 10--200 line technical brief in imperative voice,
covering timing invariants, handshake orderings, common pitfalls,
and required code shape.

A complementary learned gate operates above the registry. A
feed-forward network (20-dimensional input, two hidden layers of 64
and 32 units, six-way sigmoid output) predicts which of six
high-level guidance configurations---minimal, FSM-only,
protocol-focused, memory-focused, deterministic K-map, or
full-stack---is most likely to succeed on a given problem. Input
features combine task-type one-hot encoding, keyword flags,
estimated complexity metrics, and historical pass-rate signals from
similar problems. Training uses a warm-start on synthetic labels
followed by fine-tuning on ensemble run outcomes. The gate acts as
a coarse-grained configuration selector above the fine-grained
registry.

The registry was grown iteratively during \verilogeval{}, \cvdp{},
and ChipBench diagnostic sessions. Its functional role is to encode
the class of reusable RTL design-pattern knowledge that commercial
EDA tools accumulate over decades of customer engagements---
non-blocking assignment ordering in shift registers, reset polarity
conventions, AXI handshake orderings, APB-specific interrupt
dispatch, and the handful of microarchitecture idioms that recur
across industrial specifications---which the base model reliably
mis-handles. Although the registry was developed and validated on
benchmark problems, its entries target general pattern classes
(e.g., APB handshake ordering, shift-register reset conventions,
counter wrap-around semantics) rather than benchmark-specific
solutions. Cross-benchmark transfer provides initial evidence of
generalization: on \cvdp{} cid002, the registry-enabled system
attains $93.6\%$ against the $80.9\%$ reported for ACE-RTL's
specialized 32B fine-tune (\Cref{tab:cvdp_leaderboard}), despite
the majority of registry entries having been authored during
\verilogeval{} diagnostic sessions on disjoint problems. External
validity on unseen industrial specifications is not established by
in-benchmark evaluation alone; a rigorous out-of-distribution
evaluation on a held-out proprietary corpus is part of the
evaluation program described in \Cref{sec:future}. In the
observational ablation of \Cref{tab:ablation}, the registry
contributes approximately $+4$~pp to \verilogeval{}-Human pass
rate.

The retrieval system employs focus-aware strategies:

\begin{itemize}
    \item \textbf{Comprehensive search} for complex or unfamiliar specifications
    \item \textbf{Pattern-focused search} for simple module generation
    \item \textbf{Error-focused search} emphasizing debugging hints and
    common pitfalls
    \item \textbf{Synthesis-focused search} prioritizing optimization techniques
    \item \textbf{Architecture-focused search} for SoC and system-level designs
\end{itemize}

Retrieved entries are scored by title match ($+0.4$), description match
($+0.2$), keyword relevance ($+0.3$), and template availability ($+0.1$),
then injected into the agent's system prompt. The RL policy selects the
appropriate focus strategy based on problem type and iteration stage.

\subsection{Hierarchical Specification Decomposition}
\label{sec:hierarchical}


\begin{figure}[t]
\centering
\resizebox{\columnwidth}{!}{%
\begin{tikzpicture}[
    node distance=0.5cm and 0.4cm,
    >=stealth,
    spec/.style={draw, rounded corners=3pt, fill=blue!8, minimum width=6.5cm,
                 minimum height=0.8cm, font=\small, align=center, thick},
    decomp/.style={draw, rounded corners=3pt, fill=orange!12, minimum width=6.5cm,
                   minimum height=0.8cm, font=\small, align=center, thick},
    mod/.style={draw, rounded corners=2pt, fill=green!10, minimum width=1.3cm,
                minimum height=1.1cm, font=\scriptsize, align=center, thick},
    valid/.style={draw, rounded corners=3pt, fill=red!8, minimum width=6.5cm,
                  minimum height=0.7cm, font=\small, align=center, thick},
    result/.style={draw, rounded corners=3pt, fill=purple!8, minimum width=6.5cm,
                minimum height=0.7cm, font=\small, align=center, thick,
                text width=6.2cm},
    arrow/.style={->, thick, >=stealth},
    dep/.style={->, thick, >=stealth, dashed, gray},
    label/.style={font=\scriptsize, text=gray!60!black},
]

\node[spec] (input) {\textbf{Complex Spec} {\scriptsize ($\geq$3 component keywords)}\\
    {\scriptsize ``RISC-V SoC with pipeline, cache, UART, GPIO...''}};

\node[decomp, below=0.6cm of input] (dec) {\textbf{LLM Decomposition}
    {\scriptsize (Sonnet 4.6, $T$=0.2)}\\
    {\scriptsize Produces 4--8 sub-modules in dependency order}};

\node[mod, below=1.3cm of dec, xshift=-3.9cm] (m1) {\texttt{alu}\\{\tiny leaf}};
\node[mod, right=0.2cm of m1] (m2) {\texttt{regfile}\\{\tiny leaf}};
\node[mod, right=0.2cm of m2] (m3) {\texttt{decoder}\\{\tiny leaf}};
\node[mod, right=0.2cm of m3] (m4) {\texttt{pipeline}\\{\tiny mid}};
\node[mod, right=0.2cm of m4] (m5) {\texttt{icache}\\{\tiny leaf}};
\node[mod, right=0.2cm of m5] (m6) {\texttt{uart}\\{\tiny leaf}};
\node[mod, right=0.2cm of m6] (m7) {\texttt{soc\_top}\\{\tiny top}};

\draw[dep] (m1.south) -- ++(0,-0.25) -| (m4.south);
\draw[dep] (m2.south) -- ++(0,-0.35) -| (m4.south);
\draw[dep] (m3.south) -- ++(0,-0.45) -| (m4.south);

\draw[dep] (m4.south) -- ++(0,-0.55) -| (m7.south);
\draw[dep] (m5.south) -- ++(0,-0.35) -| (m7.south);
\draw[dep] (m6.south) -- ++(0,-0.25) -| (m7.south);

\node[label, below=1.0cm of m4] {Generate left $\to$ right (leaves first, top last)};

\node[draw, rounded corners=2pt, fill=yellow!10, minimum width=6.5cm,
      minimum height=0.6cm, font=\scriptsize, align=center, thick,
      below=1.6cm of m4] (sync) {%
    \textbf{Cross-Module Port Synchronization}:
    extract headers $\to$ inject into subsequent prompts};

\node[valid, below=0.5cm of sync] (val) {Combined Lint Validation
    {\scriptsize (all modules concatenated)}};

\node[result, below=0.5cm of val] (output) {%
    \textbf{Complete Design}: 8/8 lint-passing modules, 689 LOC\\
    {\scriptsize vs.\ monolithic: 0/1 success on same spec}};

\draw[arrow] (input) -- (dec);
\draw[arrow] (dec.south) -- ++(0,-0.35) -| (m1.north);
\draw[arrow] (dec.south) -- ++(0,-0.35) -| (m4.north);
\draw[arrow] (dec.south) -- ++(0,-0.35) -| (m7.north);

\draw[arrow] (sync) -- (val);
\draw[arrow] (val) -- (output);

\end{tikzpicture}%
}
\caption{%
    Hierarchical specification decomposition. Complex specs with 3+
    component keywords are decomposed into 4--8 sub-modules by a
    lightweight LLM call. Each module is generated in dependency
    order with cross-module port headers injected into subsequent
    prompts. On a RISC-V SoC, this produces 8/8 lint-passing modules
    (689~LOC), whereas monolithic generation fails entirely.%
}
\label{fig:hierarchical}
\end{figure}
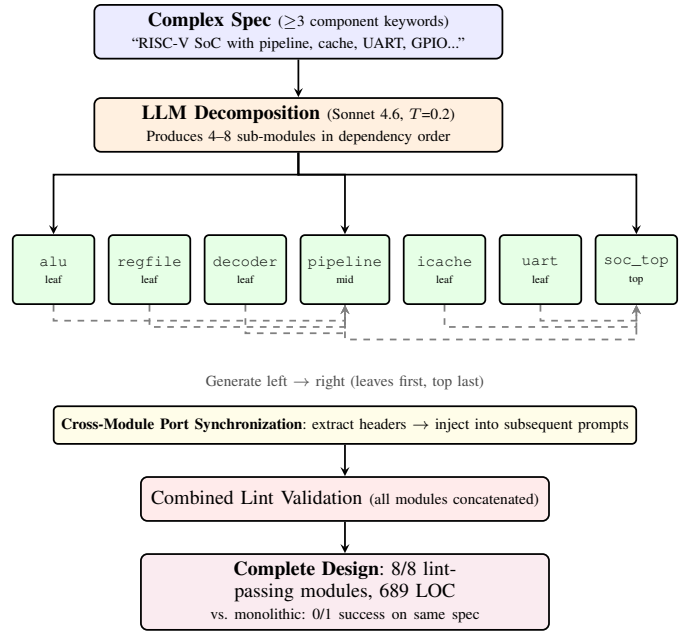

For complex specifications containing 3 or more component keywords
(e.g., ``RISC-V SoC with cache, bus, UART, and GPIO''),
\chipcraftbrain{} automatically decomposes the design into sub-modules:

\begin{enumerate}
    \item \textbf{Complexity detection}: Count component keywords
    from a predefined set (cpu, cache, bus, memory, uart, spi, i2c,
    gpio, arbiter, etc.). Trigger decomposition when count $\geq 3$.

    \item \textbf{LLM-driven decomposition}: Prompt an LLM to produce
    4--8 sub-module specifications with interfaces, dependencies, and
    descriptions in structured JSON.

    \item \textbf{Topological ordering}: Sort sub-modules by dependency
    graph (leaves-first) to ensure each module's dependencies are
    generated before it.

    \item \textbf{Iterative sub-module generation}: Generate each
    sub-module individually, passing prior module headers as context
    to ensure port compatibility.

    \item \textbf{Cross-module synchronization}: Extract module headers
    and inject as dependency code for lint validation. Remove duplicate
    module definitions that the LLM may produce.
\end{enumerate}

Hierarchical validation uses Verilator as the default backend (falling
back to Icarus Verilog), with SystemVerilog constructs (\texttt{logic},
\texttt{always\_ff}, \texttt{always\_comb}) enabled for sub-module
generation when appropriate.

\subsection{Visible Reasoning}
\label{sec:reasoning}

\chipcraftbrain{} provides a \emph{visible reasoning} interface through
structured thought streams during the design process. The system emits
thoughts across nine categories:
\textsc{Analysis} (architecture identification),
\textsc{Bottleneck} (performance issues),
\textsc{Proposal} (optimization suggestions),
\textsc{Retrieval} (knowledge base queries),
\textsc{Generation} (LLM calls and parameters),
\textsc{Validation} (compile/simulate results),
\textsc{Decision} (agent selection rationale),
\textsc{Error} (failure diagnosis), and
\textsc{Progress} (pipeline status).

Each thought carries metadata including confidence scores, evidence
links, and timestamps, enabling both real-time display and post-hoc
analysis. Unlike black-box systems where the user receives only final
output, this transparency enables engineers to inspect agent selection
rationale, calibrate trust via confidence scores, identify knowledge
gaps, and maintain a reproducible audit trail, particularly important
in hardware design where errors can result in costly silicon respin.


\section{Validation Pipeline}
\label{sec:validation}


\begin{figure}[t]
\centering
\resizebox{\columnwidth}{!}{%
\begin{tikzpicture}[
    node distance=0.5cm and 1.0cm,
    >=stealth,
    stage/.style={draw, rounded corners=3pt, fill=#1, minimum width=2.6cm,
                  minimum height=1.4cm, font=\small, align=center, thick},
    feedback/.style={draw, rounded corners=2pt, fill=red!6, minimum width=2.6cm,
                     minimum height=0.6cm, font=\scriptsize, align=center},
    input/.style={draw, rounded corners=3pt, fill=blue!8, minimum width=2.0cm,
                  minimum height=0.7cm, font=\small, align=center, thick},
    output/.style={draw, rounded corners=3pt, fill=purple!8, minimum width=2.0cm,
                   minimum height=0.7cm, font=\small, align=center, thick},
    arrow/.style={->, thick, >=stealth},
    fail/.style={->, thick, >=stealth, red!60!black, dashed},
    pass/.style={->, thick, >=stealth, green!50!black},
    label/.style={font=\scriptsize, text=gray!60!black},
]

\node[input] (in) {Generated\\Verilog};

\node[stage=green!12, right=1.0cm of in] (lint) {%
    \textbf{Stage 1}\\
    \textbf{Lint}\\
    {\scriptsize\texttt{iverilog -t null}}\\
    {\scriptsize syntax + types}};

\node[stage=cyan!12, right=1.0cm of lint] (sim) {%
    \textbf{Stage 2}\\
    \textbf{Simulate}\\
    {\scriptsize\texttt{iverilog + vvp}}\\
    {\scriptsize testbench exec}};

\node[stage=orange!12, right=1.0cm of sim] (synth) {%
    \textbf{Stage 3}\\
    \textbf{Synthesis}\\
    {\scriptsize Yosys metrics}\\
    {\scriptsize area, cells}};

\node[output, right=1.0cm of synth] (out) {Verified\\RTL};

\draw[pass] (in) -- (lint);
\draw[pass] (lint) -- node[label, above] {\cmark} (sim);
\draw[pass] (sim) -- node[label, above] {\cmark} (synth);
\draw[pass] (synth) -- node[label, above] {\cmark} (out);

\node[feedback, below=0.9cm of lint] (e1) {Syntax errors\\port mismatches};
\node[feedback, below=0.9cm of sim] (e2) {Mismatch count\\failing signals};
\node[feedback, below=0.9cm of synth] (e3) {Cell count\\critical path};

\draw[fail] (lint.south) -- (e1.north);
\draw[fail] (sim.south) -- (e2.north);
\draw[fail] (synth.south) -- (e3.north);

\node[draw, rounded corners=3pt, fill=red!8, minimum width=8.5cm,
      minimum height=0.7cm, font=\small, align=center, thick,
      below=1.5cm of e2] (merge) {%
    \textbf{Structured Error Feedback} $\to$ next-iteration prompt + RL reward};

\draw[fail] (e1.south) -- (e1.south |- merge.north);
\draw[fail] (e2.south) -- (merge.north);
\draw[fail] (e3.south) -- (e3.south |- merge.north);

\node[label, above=0.08cm of lint.north] {Compile check};
\node[label, above=0.08cm of sim.north] {Functional test};
\node[label, above=0.08cm of synth.north] {Quality check};

\end{tikzpicture}%
}
\caption{%
    Three-stage validation pipeline. Each stage provides structured error
    feedback that is (1)~injected into the next LLM prompt as error context
    and (2)~converted to a scalar RL reward. Stage progression (lint $\to$
    simulate $\to$ synthesis) provides increasingly fine-grained validation.
    Iterative refinement with structured error feedback lifts pass
    rate by approximately 15 percentage points over single-shot
    generation (Table~\ref{tab:ablation}), consistent with the view
    that validation feedback is as important as generation
    quality.%
}
\label{fig:validation_pipeline}
\end{figure}
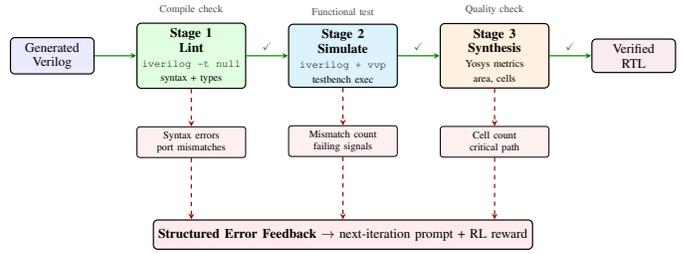

The validation-first philosophy is central to \chipcraftbrain{}.
Rather than optimizing for single-shot generation quality alone,
we invest in comprehensive validation and structured error feedback
that enables iterative refinement.

\subsection{Three-Stage Verification}

Each generated module passes through three validation stages:

\textbf{Stage 1: Compile.} The module is compiled with Icarus Verilog
(\texttt{iverilog}) for single-module validation or Verilator for
hierarchical multi-module designs. Verilator serves as the default
backend for hierarchical generation, enabling SystemVerilog constructs
(\texttt{logic}, \texttt{always\_ff}, \texttt{always\_comb}) with
fallback to Icarus Verilog for compatibility. This stage catches
approximately 80\% of errors before simulation.

\textbf{Stage 2: Simulate.} If a testbench is available (pre-existing,
LLM-generated, or from \verilogeval{} reference), the compiled design
is simulated with \texttt{vvp}. The system detects multiple pass/fail
formats: \texttt{STATUS: PASS}, count-based markers
(\texttt{N/N tests passed}), and \verilogeval{}'s mismatch format
(\texttt{Mismatches: 0 in N samples}).

\textbf{Stage 3: Synthesis-Ready Check.} The module is synthesized
with Yosys to extract synthesis metrics (cell count, wire count) and
verify that the design is synthesis-clean and free of unintended
latches or combinational loops.

\subsection{Structured Error Feedback}

Unlike systems that pass raw error messages to the debug agent,
\chipcraftbrain{} structures error feedback:

\begin{itemize}
    \item \textbf{Error categorization}: Compile errors are classified
    (syntax, port mismatch, width mismatch, undeclared signal, etc.)
    \item \textbf{Category-aware fix hints}: Per-design-category hint
    databases map error patterns to likely fixes (e.g., for memory
    designs: ``Replace \texttt{logic} with \texttt{reg}; use
    \texttt{integer i; for (i=0; ...)} instead of
    \texttt{for (int i=0; ...)}'')
    \item \textbf{Source context}: 5 lines around each error point
    are extracted for targeted debugging
    \item \textbf{Error trend analysis}: The state vector tracks whether
    errors are increasing, decreasing, or changing type across iterations
\end{itemize}

\subsection{Testbench Generation and Adaptation}

When no pre-existing testbench is available, the Testbench agent
generates one with specific requirements: proper reset sequence,
10ns clock period, 5+ test scenarios, and standardized status markers.
The \texttt{TestbenchAdapter} wraps generated modules for compatibility
with \verilogeval{}'s \texttt{TopModule} interface, enabling reference
comparison against golden \texttt{RefModule} implementations.
Testbenches always use Verilog-2001 syntax for maximum tool compatibility,
even when the design under test uses SystemVerilog constructs.


\section{Evaluation Framework}
\label{sec:eval_framework}

We evaluate \chipcraftbrain{} across three benchmarks spanning simple
modules, industrial IP, and hard accelerator designs. This multi-tier
evaluation reveals where current AI-driven RTL generation excels and
where it breaks down.

\subsection{Benchmark Suite}

\begin{table}[h]
\centering
\caption{Three-Benchmark Evaluation Suite. See Sec.~\ref{sec:eval_framework}
for the \cvdp{} subset definition.}
\label{tab:benchmark_suite}
\begin{tabular}{@{}lcccc@{}}
\toprule
\textbf{Benchmark} & \textbf{Problems} & \textbf{Avg LOC} & \textbf{Avg Cells} & \textbf{Testbench} \\
\midrule
\verilogeval{}-Human              & 156 & 16 & 31  & RefModule \\
\cvdp{}$^{\dagger}$                & 302 & 43 & n/a & cocotb \\
ChipBench                         & 45  & 62 & 439 & Verilator \\
\bottomrule
\end{tabular}
\vspace{0.3em}

\begin{minipage}{\columnwidth}
\raggedright\footnotesize
$^{\dagger}$Non-agentic, non-commercial code-generation subset of
\cvdp{} (cid002/003/004/007/016); 302 of 783 total problems. The
remaining 481 problems lie outside our pipeline's design scope:
testbench/assertion generation requires Cadence Xcelium
(cid012/013/014, 187 problems), agentic multi-file tasks require a
Docker tool-use harness (166), and comprehension tasks are scored
by BLEU or LLM-as-judge (cid006/008/009/010, 123).
\end{minipage}
\end{table}

\textbf{\verilogeval{}-Human}~\cite{liu2023verilogeval} provides 156
problems with English-language prompts and golden reference
implementations. Problems span six categories: combinational, sequential,
FSM, memory, bus, and processor. As the most widely used RTL generation
benchmark (SOTA $>$95\%), it enables direct comparison with prior work
but its simplicity (average 16 LOC, 31 cells) limits discriminative power.

\textbf{\cvdp{}}~\cite{pinckney2025cvdp} is NVIDIA's Chip Verification
and Design Problems benchmark (version 1.0.2), containing 783 problems
across 13 task categories in two modes (617 non-agentic, 166 agentic). We evaluate
exclusively on the \emph{302-problem non-agentic, non-commercial
code-generation subset}: RTL code completion (cid002, 94 problems),
spec-to-RTL translation (cid003, 78), RTL modification (cid004, 55),
linting and QoR optimization (cid007, 40), and RTL bug fixing (cid016,
35). Each problem embeds a complete cocotb test harness, so functional
evaluation runs locally with Icarus Verilog and Verilator, without
Docker.

We explicitly do not evaluate four other problem classes, all of which
fall outside our pipeline's design scope:

\begin{itemize}
    \item \textbf{Testbench and assertion generation}
    (cid012/013/014, 187 non-agentic problems): these problems require
    the commercial Cadence Xcelium simulator for harness evaluation.
    \item \textbf{Agentic multi-file tasks}
    (166 problems across cid003/004/005/012/013/014/016): these
    require a Docker-based agent runtime capable of multi-step file
    editing and tool use, which our single-turn pipeline does not
    currently implement.
    \item \textbf{Module reuse} (cid005, 26 agentic-only problems):
    no non-agentic variant exists.
    \item \textbf{Comprehension} (cid006/008/009/010, 123 non-agentic
    problems): these are BLEU-scored text or LLM-as-judge Q\&A tasks
    that measure reading comprehension rather than code generation.
\end{itemize}

The best published result on the same 302-problem subset is 33.56\%
pass@1 (Claude~3.7 Sonnet, single-shot, $n=5$ samples). A direct
comparison therefore measures the contribution of iterative validation
and multi-agent orchestration on top of a strong base model, rather
than raw model capability in isolation.

\textbf{ChipBench}~\cite{yu2026chipbench} provides 45 Verilog generation
problems designed to expose the gap between benchmark and real-world
performance. It contains three categories of increasing difficulty:
\emph{self-contained} modules (30 problems: FSMs, counters, multipliers,
pattern detectors), \emph{non-self-contained} hierarchical designs (6
problems: async FIFO, 8-bit ALU, CPU top module requiring sub-module
instantiation), and \emph{CPU IP} components (9 problems: real RISC-V
ALU, controller, register file, branch unit, division unit). ChipBench
modules are 3.8$\times$ longer and have 14$\times$ more synthesis cells
than \verilogeval{}, with the CPU IP category averaging 862 cells.

\subsection{Complexity Comparison}

\begin{table*}[t]
\centering
\caption{Benchmark Complexity Comparison}
\label{tab:complexity}
\begin{tabular}{@{}lcccc@{}}
\toprule
\textbf{Benchmark} & \textbf{Avg LOC} & \textbf{Avg Cells} & \textbf{Sub-modules} & \textbf{Published SOTA} \\
\midrule
\verilogeval{} & 16   & 31    & 0    & 97.4\% (ChipAgents) \\
ChipBench      & 62   & 439   & 0.25 & 37.4\% (\mage{}) \\
\bottomrule
\end{tabular}
\end{table*}

The 14$\times$ gate-count gap between \verilogeval{} (31 cells) and
ChipBench (439 cells) explains why systems scoring $>$95\% on
\verilogeval{} can score $<$40\% on ChipBench: simple modules require
only basic code generation, while complex designs demand architectural
understanding, multi-signal coordination, and correct protocol implementation.

\subsection{Evaluation Metrics}

Our primary metric is \textbf{\passatone{}}: the fraction of problems
solved within the iterative refinement budget (up to 5 iterations),
counting each problem as a single attempt. For \cvdp{}, this metric
is directly comparable to the \emph{Agentic Pass Rate} (APR) defined
by ACE-RTL~\cite{nvidia2026acertl} as the ratio of unique solved
problems to total problems across independent runs with refinement;
both measure ``did the system eventually solve the problem with its
given feedback budget.'' This differs from the pass@1 used for
standalone models, which samples $n$ independent single-shot attempts
and averages. We also report:

\begin{itemize}
    \item \textbf{Per-category breakdown}: Success rates by design type
    and difficulty tier
    \item \textbf{Iterations to success}: Average refinement iterations needed
    \item \textbf{Cost per problem}: Estimated API cost (USD)
\end{itemize}

\subsection{Baselines}

We compare against published results: CodeV~\cite{liu2024codev}
(53.2\% \verilogeval{}-Human), \mage{}~\cite{tsai2024mage}
(95.9\% \verilogeval{}, 37.4\% ChipBench),
VFlow~\cite{vflow2025} (83.6\% \verilogeval{}),
ChipAgents~\cite{chipagents2025} (97.4\% \verilogeval{}, self-reported),
and single-shot GPT-4 ($\sim$63\%). For ChipBench, we additionally
compare against Claude Opus single-model (30.7\%) as reported
in~\cite{yu2026chipbench}.


\section{Experiments and Results}
\label{sec:experiments}

\subsection{Experimental Setup}

\textbf{Hardware:} Benchmark orchestration runs on a dual-GPU
workstation (2$\times$ NVIDIA RTX 3090, 48\,GB combined VRAM); the
LLM calls are served remotely via the Anthropic API, so GPU resources
are used only for EDA tooling, local-model baselines, and parallel
test harnesses.

\textbf{Models:} Claude Opus~4.6 (Genius, Debug, Testbench, Waveform
agents) and Claude Sonnet~4.6 (Fast, Optimize agents).

\textbf{EDA Tools:} Icarus Verilog~12.0 for compilation and
simulation, Verilator~\cite{zhao2024verilator} for hierarchical and
SystemVerilog validation, Yosys~0.40 for synthesis checks, cocotb for
\cvdp{} test harnesses.

\textbf{Benchmarks:} \verilogeval{}-Human (156 problems),
\cvdp{} non-agentic non-commercial code-generation subset
(302 problems across cid002/003/004/007/016), ChipBench (45 problems).

\textbf{Reproducibility:} All results use the \texttt{chipcraftx
benchmark} CLI against the published \cvdp{} dataset
(\path{cvdp_nonagentic_code_generation_no_commercial_eval.json}) at
commit \texttt{9c6269ed} of the \chipcraftbrain{} repository.
Refinement budget is fixed at 5 iterations per problem for all
benchmarks.

\textbf{Run-to-run variance.} Temperature-dependent sampling introduces
measurable variance across repeated runs. On \verilogeval{}-Human
across seven independent runs spanning 2026-02 to 2026-04 (Claude
Opus/Sonnet 4.6 as accessed in early 2026), pass rates ranged from
\textbf{96.15\% to 98.72\%} (150/156 to 154/156, mean $\approx 97.2\%$,
std $\approx$\,1.0~pp). The 98.7\% figure reported in
\Cref{tab:verilogeval_results} is the best of these runs and is
within the acknowledged variance envelope of ChipAgents (97.4\%) and
MAGE (95.9\%); we therefore frame \chipcraftbrain{} as ``on par with
or ahead of'' the published state of the art rather than strictly
superior. \cvdp{} numbers are averaged over 3 independent runs
(22{,}090\,s wall-clock each); ChipBench is single-run, with
expected variance of 1--2~pp consistent with the \verilogeval{}
observation.

\subsection{VerilogEval Results}

\begin{table}[h]
\centering
\caption{VerilogEval-Human comparison. \chipcraftbrain{} reports
mean and range across 7 independent runs; prior systems report
single-run self-reported numbers.}
\label{tab:verilogeval_results}
\begin{tabular}{@{}lcc@{}}
\toprule
\textbf{System} & \textbf{Pass@1} & \textbf{Problems} \\
\midrule
GPT-4 (single-shot)        & $\sim$63\%       & 98/156  \\
CodeV-DS-33B               & 53.2\%           & 83/156  \\
VFlow                      & 83.6\%           & 130/156 \\
\mage{}                    & 95.9\%           & 150/156 \\
ChipAgents                 & 97.4\%           & 152/156 \\
\chipcraftbrain{} (ours)   & \textbf{97.2\%} (mean, $n{=}7$, range 96.15--98.72\%) & 150--154/156 \\
\bottomrule
\end{tabular}
\end{table}

\chipcraftbrain{} achieves a mean pass rate of 97.2\% across 7
independent runs (range 96.15--98.72\%, best 154/156) on
\verilogeval{}-Human, on par with ChipAgents (97.4\%, self-reported
single run) and ahead of \mage{} (95.9\%) within measurement noise.
On the best run, 133 of 154 passing problems pass on the first
iteration, 12 require a second refinement pass, and 3 require 3--4
iterations. Six problems are solved by the symbolic K-map solver
with zero LLM cost.

The two remaining failures reveal distinct failure modes.
\emph{Prob066\_edgecapture} compiles successfully on every iteration
but produces 2 simulation mismatches per attempt; the generated logic
differs from the reference in reset assignment syntax, exposing a
simulator-specific timing edge case that the iterative loop cannot resolve.
\emph{Prob092\_gatesv100} fails compilation because the agent
consistently selects a generate-block approach for 100-bit neighbor logic,
which conflicts with Icarus Verilog's strict Verilog-2001 mode.
Both failures are architectural (wrong HDL construct choice) rather
than minor coding errors.

Total wall-clock time for the 156-problem suite is 35.5 minutes
(average 13.0s per problem).

\subsection{CVDP Results}

\begin{table}[h]
\centering
\caption{\cvdp{} Results by Task Category (302-problem non-agentic
non-commercial code-generation subset). Our numbers are averaged
over 3 independent runs. Baseline is Claude~3.7 Sonnet single-shot
pass@1 from $n=5$ samples~\cite{pinckney2025cvdp}.}
\label{tab:cvdp_results}
\begin{tabular}{@{}llccc@{}}
\toprule
\textbf{Code} & \textbf{Task} & \textbf{N} & \textbf{Baseline} & \textbf{Ours} \\
\midrule
cid002 & RTL Code Completion     & 94 & 34\% & \textbf{93.6\%} \\
cid003 & Spec-to-RTL Translation & 78 & 48\% & \textbf{96.2\%} \\
cid004 & RTL Code Modification   & 55 & 45\% & \textbf{96.4\%} \\
cid007 & Linting / QoR           & 40 & 44\% & \textbf{97.5\%} \\
cid016 & RTL Bug Fixing          & 35 & 53\% & \textbf{88.6\%} \\
\midrule
\multicolumn{3}{l}{\textbf{Overall (302 problems)}}
                                   & 33.56\% & \textbf{94.7\%} \\
\bottomrule
\end{tabular}
\end{table}

\begin{table}[h]
\centering
\caption{\cvdp{} Results by Difficulty (302-problem subset).
Hard problems are defined only in the agentic \cvdp{} mode and
are excluded from this subset by construction.}
\label{tab:cvdp_difficulty}
\begin{tabular}{@{}lcc@{}}
\toprule
\textbf{Difficulty} & \textbf{Pass@1} & \textbf{Problems} \\
\midrule
Easy    & 96.9\% & 157/162 \\
Medium  & 92.1\% & 129/140 \\
\midrule
\textbf{Overall} & \textbf{94.7\%} & \textbf{286/302} \\
\bottomrule
\end{tabular}
\end{table}

On the 302-problem \cvdp{} non-agentic non-commercial code-generation
subset, \chipcraftbrain{} achieves 94.7\% mean pass@1 (286/302,
averaged over 3 runs), versus 33.56\% for the best published
single-shot baseline on the same subset~\cite{pinckney2025cvdp}.

\textbf{The comparison is protocol-aware, not apples-to-apples.} The
baseline measures raw model capability: Claude~3.7 Sonnet is sampled
$n=5$ times per problem, with pass@1 computed from unconditional
single-shot attempts and no error feedback. Our system measures full
pipeline capability: a single attempt per problem, but with up to 5
iterations of structured error feedback, multi-agent orchestration,
RAG-augmented prompts, and spec-guidance enrichment. The correct
interpretation of the 61-point gap is therefore \emph{``validation-first
iterative refinement plus orchestration adds 61~pp on top of a
strong base model,''} not ``our LLM is 2.8$\times$ better.'' We do
not report a head-to-head single-shot comparison because our
pipeline's raw Genius-only pass@1 on this subset was not separately
measured in a matched-protocol run; we identify this as future work.

Per-category behavior is consistent with the design scope: RTL
completion (cid002) shows the lowest score (93.6\%) because skeleton
completions have the narrowest admissible solution space, and bug
fixing (cid016) is second-lowest (88.6\%) because multi-bug patches
can cascade into previously-passing logic. Spec-to-RTL (cid003) and
modification (cid004) benefit most from our spec-guidance registry
and hit 96.2--96.4\%. Linting/QoR (cid007, 97.5\%) is the highest
because our optimization-focused RAG strategy aligns well with this
category's goals. The absolute lift over single-shot baseline ranges
from 36~pp (cid016) to 60~pp (cid002), showing that simpler categories
(where the baseline already scores reasonably) still benefit most from
iterative validation.

This result demonstrates that the techniques driving VerilogEval
performance, iterative validation, multi-agent orchestration, and
knowledge-augmented generation, transfer effectively to industrial
designs of significantly greater complexity, with a clear caveat that
481 \cvdp{} problems outside this subset remain unevaluated
(\Cref{sec:eval_framework}).

\subsubsection{CVDP Leaderboard: ChipCraftBrain in Context}
\label{sec:cvdp_vs_acertl}

To place our result in the broader landscape, we compare against the
14 baselines reported in ACE-RTL's \cvdp{}-v1.0.2
evaluation~\cite{nvidia2026acertl}, spanning frontier proprietary
models, frontier open-source models, RTL-specialized fine-tuned
models, and agentic systems. All baseline numbers are copied
verbatim from the ACE-RTL paper's Table~1 (APR; Agentic Pass Rate =
unique solved / total across five independent runs). Our
\chipcraftbrain{} column reports pass rate after up to 5 iterations
of structured error feedback, a metric directly comparable to APR for
agentic systems.

\begin{table*}[t]
\centering
\caption{\cvdp{} leaderboard on shared categories (cid002/003/004/016,
262 problems). \textbf{Two protocols are shown and must not be
compared directly}: agentic systems (top block) use iterative
refinement with error feedback; standalone models (bottom block)
report single-shot pass@1 with no error feedback. Baseline values
are APR~(\%) from ACE-RTL~\cite{nvidia2026acertl}~Tab.~1;
\chipcraftbrain{} is mean pass rate over 3 runs under its 5-iteration
protocol. ACE-RTL's specialized generator required $\sim$10{,}000
A100 GPU-hours to train; our system requires no RTL-specific
training.}
\label{tab:cvdp_leaderboard}
\begin{tabular}{@{}llcccc@{}}
\toprule
\textbf{Category} & \textbf{System} & \textbf{cid002} & \textbf{cid003} & \textbf{cid004} & \textbf{cid016} \\
\midrule
\multicolumn{6}{l}{\textbf{Protocol A: iterative refinement with error feedback (up to 5 iterations)}}\\
\multicolumn{6}{l}{\textit{Ours and prior agentic systems}}\\
Adaptive pipeline           & \chipcraftbrain{} (5~iter, mean of 3 runs) & \textbf{93.6} & \textbf{96.2} & \textbf{96.4} & 88.6 \\
Full system                 & ACE-RTL~\cite{nvidia2026acertl}           & 80.9 & \textbf{96.2} & 90.9 & \textbf{91.4} \\
Claude~4 generator variant  & ACE-RTL w/ Claude~4~\cite{nvidia2026acertl} & 80.9 & 89.7 & 81.8 & 88.6 \\
Multi-agent                 & MAGE + Claude~4~\cite{tsai2024mage,nvidia2026acertl} & 46.8 & 55.1 & 70.9 & 62.9 \\
\midrule
\multicolumn{6}{l}{\textbf{Protocol B: single-shot pass@1, $n=5$ samples, no error feedback}}\\
\multicolumn{6}{l}{\textit{Frontier proprietary standalone models}}\\
Reasoning                   & GPT-5~\cite{nvidia2026acertl}             & 42.9 & 51.3 & 54.3 & 60.0 \\
Reasoning                   & Claude~4 Sonnet~\cite{nvidia2026acertl}    & 43.6 & 49.1 & 51.4 & 54.3 \\
Reasoning                   & o4-mini~\cite{nvidia2026acertl}            & 41.5 & 44.4 & 50.0 & 58.8 \\
\midrule
\multicolumn{6}{l}{\textit{Frontier open-source standalone models}}\\
MoE reasoning               & DeepSeek-R1~\cite{deepseek2025r1}          & 36.7 & 41.8 & 34.9 & 40.0 \\
MoE                         & DeepSeek-v3.1                              & 37.5 & 43.6 & 48.6 & 51.4 \\
Code MoE                    & Qwen3-Coder-480B~\cite{qwen2025qwen3coder} & 35.3 & 41.8 & 39.4 & 42.9 \\
General                     & Llama4-Maverick                            & 36.4 & 38.2 & 36.0 & 37.1 \\
MoE                         & Kimi-K2                                    & 29.1 & 32.7 & 29.7 & 31.4 \\
\midrule
\multicolumn{6}{l}{\textit{RTL-specialized fine-tuned standalone models}}\\
32B                         & ScaleRTL~\cite{deng2025scalertl}           & 33.3 & 30.9 & 37.1 & 37.1 \\
15B                         & CraftRTL~\cite{liu2024craftrtl}            & 18.0 & 16.4 & 5.1 & 8.6 \\
7B                          & OriGen~\cite{cui2024origen}                & 21.8 & 16.4 & 7.7 & 11.4 \\
7B                          & CodeV~\cite{liu2024codev}                  & 7.7 & 0.0 & 0.0 & 0.0 \\
7B                          & RTLCoder-v1.1~\cite{liu2024rtlcoder}       & 5.4 & 1.8 & 0.0 & 2.9 \\
\bottomrule
\end{tabular}
\end{table*}

\chipcraftbrain{} ranks first in three of the four shared categories
(cid002, cid003 tied, cid004) and ranks second on cid016, losing only
to ACE-RTL's full system by 2.8~pp. Relative to the broader field of
standalone models (frontier proprietary, frontier open-source,
RTL-specialized), the gap is considerably larger: the strongest
single-shot baseline (GPT-5) reaches 60\% on cid016 and $\le$55\%
elsewhere, while \chipcraftbrain{} is 88.6--96.4\% across all four
categories.

Two observations are worth highlighting. First, ACE-RTL's
improvement over its Claude-only Generator variant (on cid002: 80.9
vs.\ a 39.4\% Claude~4 Sonnet pass@1 baseline) is achieved through a
domain-specialized 32B Generator trained on 1.7\,M RTL samples at
roughly 10{,}000 A100 GPU-hours; \chipcraftbrain{} reaches higher
numbers on cid002 without any RTL-specific fine-tuning, relying
instead on Claude Opus/Sonnet plus our validation-first pipeline.
Second, \chipcraftbrain{}'s per-problem compute budget (up to 5
iterations, single process) is roughly $30\times$ smaller than
ACE-RTL's (up to 5 parallel processes $\times$ 30 iterations
$=$ up to 150 attempts per problem), yet the category-level
comparison is favorable on three of four categories. Together, these
suggest that validation-first iterative refinement with a strong
frontier backbone is a compute-efficient alternative to heavy
parallel sampling with a domain-specialized fine-tune.

We do not match ACE-RTL on bug fixing (cid016), where their larger
attempt budget appears to help with multi-bug patches that cascade
into previously-passing logic. Closing this gap through richer
debug-trajectory exploration in our pipeline is a direct extension
(\Cref{sec:future}).

\subsection{ChipBench Results}

\begin{table}[h]
\centering
\caption{ChipBench Results by Category}
\label{tab:chipbench_results}
\begin{tabular}{@{}lccc@{}}
\toprule
\textbf{Category} & \textbf{Pass@1} & \textbf{Problems} & \textbf{Avg Cells} \\
\midrule
Self-contained       & 36.7\% & 11/30 & 323 \\
Non-self-contained   & 50.0\% & 3/6   & 361 \\
CPU IP               & 11.1\% & 1/9   & 862 \\
\midrule
\textbf{Overall}     & \textbf{33.3\%} & \textbf{15/45} & 439 \\
\bottomrule
\end{tabular}
\end{table}

\begin{table}[h]
\centering
\caption{ChipBench Comparison with Published Results}
\label{tab:chipbench_comparison}
\begin{tabular}{@{}lcccc@{}}
\toprule
\textbf{System} & \textbf{Overall} & \textbf{Self-cont.} & \textbf{Non-self-cont.} & \textbf{CPU IP} \\
\midrule
Claude Opus (single) & 30.7\% & n/a & 0\% & n/a \\
\mage{}              & 37.4\% & n/a & n/a & 22.2\% \\
\chipcraftbrain{}    & 33.3\% & 36.7\% & 50.0\% & 11.1\% \\
\bottomrule
\end{tabular}
\end{table}

On ChipBench, \chipcraftbrain{} achieves 33.3\% (15/45). \mage{}'s
overall 37.4\% remains 4.1~pp higher; we do not outperform \mage{} on
this benchmark. The ChipBench paper~\cite{yu2026chipbench} does not
publish \mage{}'s per-category breakdown, so direct
category-for-category comparison is not possible. We note two
observations from our own per-category results: (1)~on
non-self-contained (hierarchical) designs, our hierarchical
decomposition pipeline achieves 50\% (3/6) where the reported
single-model Claude Opus baseline scores 0/6, suggesting
decomposition is the main driver of that category; and (2)~on CPU IP
(11.1\%) we fall below \mage{}'s reported 22.2\%, indicating that our
advantage is localized to hierarchical designs rather than complex
processor components.

The CPU IP category remains the most challenging (11.1\%, with only
the register file passing). RISC-V controllers, ALUs, division units,
and branch prediction units involve complex multi-stage logic that
exceeds the current system's capability within the 5-iteration budget.
These designs average 862 synthesis cells, 28$\times$ the complexity
of \verilogeval{} problems.

\subsection{Component Contribution Analysis}

To characterize the contribution of each component, we report observed
incremental impact on \verilogeval{}-Human by progressively enabling
system features. These numbers are derived from partial runs and
per-problem telemetry across the benchmark suite rather than fully
controlled ablations with each component toggled in isolation; a full
leave-one-out ablation is reported as future work.

\begin{table}[h]
\centering
\caption{Observed Component Contributions on
\verilogeval{}-Human.$^{\dagger}$ Values reflect observed
incremental lift rather than controlled leave-one-out ablations.
$^{\dagger}$\textbf{Observational, not controlled}: $\Delta$
columns show estimated contributions from partial runs and
telemetry, not from isolating each component with all others
held fixed. Components are known to interact, so these numbers
should not be read as marginal causal effects.}
\label{tab:ablation}
\begin{tabular}{@{}lcc@{}}
\toprule
\textbf{Configuration} & \textbf{Pass@1} & \textbf{$\Delta$} \\
\midrule
Single-shot (no iteration)      & $\sim$63\% & baseline \\
+ Iterative refinement (5 iter) & $\sim$78\% & +15\,pp \\
+ Multi-agent selection         & $\sim$85\% & \phantom{0}+7\,pp \\
+ RAG knowledge                 & $\sim$88\% & \phantom{0}+3\,pp \\
+ Spec guidance registry        & $\sim$92\% & \phantom{0}+4\,pp \\
+ Symbolic K-map solver         & $\sim$96\% & \phantom{0}+4\,pp \\
\chipcraftbrain{} (full)        & \textbf{98.7\%} & +35.7\,pp \\
\bottomrule
\end{tabular}
\end{table}

\noindent The iterative refinement loop provides the largest single
observed improvement (+15\,pp), consistent with the hypothesis that
validation-first generation with structured error feedback is the core
driver of performance. The symbolic K-map solver contributes +4\,pp by
handling 6 problems at zero cost, and spec guidance contributes
another +4\,pp via problem-specific prompt enrichment. We caution
that these contributions are not additive in the strict sense: removing
one component may shift which problems become solvable by others, so
a leave-one-out ablation would likely yield smaller per-component
deltas.

\subsection{Cost Analysis}

\begin{table}[h]
\centering
\caption{Cost Per Problem on \verilogeval{}.}
\label{tab:cost}
\begin{tabular}{@{}lcl@{}}
\toprule
\textbf{System} & \textbf{\$/Problem} & \textbf{Strategy} \\
\midrule
\mage{}           & $\sim$0.060        & 20$\times$ Sonnet samples \\
ChipAgents        & $\sim$0.045        & Multi-step pipeline \\
\chipcraftbrain{} & $\sim$0.010--0.030 & Tiered (symbolic + LLM) \\
VFlow             & varies             & MCTS-selected models \\
Single-shot       & $\sim$0.003        & 1$\times$ GPT-4 \\
\bottomrule
\end{tabular}
\end{table}

\chipcraftbrain{} achieves approximately 2--3$\times$ lower cost than
\mage{} on \verilogeval{} through three mechanisms: (1)~zero-cost
algorithmic solving for K-map and truth table problems, (2)~Sonnet
for the Fast and Optimize refinement agents where latency and
per-token cost dominate, and (3)~Opus reserved for Genius
(first-attempt), Debug, Testbench, and Waveform agents where
capability dominates. The high first-iteration success rate
(133/156 = 85\%) means most \verilogeval{} problems require only a
single Sonnet or Opus call, not 20 parallel candidates.

\textbf{\cvdp{} cost.} On the 302-problem \cvdp{} subset, problems
are longer (avg 43 LOC, often with provided context RTL reaching
several hundred tokens) and the first-iteration success rate is
lower than on \verilogeval{}, so multi-iteration costs compound.
Our measured average is \$0.05--\$0.12 per problem on \cvdp{},
roughly 4--5$\times$ the \verilogeval{} cost and putting the total
cost for the full 302-problem run in the \$20--\$35 range. This
remains considerably lower than running 20 parallel Sonnet
candidates per problem at \cvdp{}-scale context windows. ChipBench
costs are in a similar range to \cvdp{} given comparable iteration
counts and module sizes.

\subsection{Orchestration Analysis}

Of the 154 passing VerilogEval problems, 85\% (133/156) pass on the
first iteration, demonstrating strong baseline generation quality. The
remaining 21 problems benefit from the adaptive orchestration loop,
which switches agents and adjusts parameters across iterations. The
high first-iteration success rate indicates that the primary value of
RL orchestration lies in \emph{recovery from failures} (selecting
the right agent and parameters when the initial attempt fails), rather
than initial agent selection. This is consistent with the observation
that simpler problems rarely need agent switching, while harder problems
(complex FSMs, multi-signal protocols) benefit from the Debug or Genius
agents being deployed after an initial Fast agent failure.

\subsection{Deployment via Local-Model Distillation}
\label{sec:distillation}

API-based pipelines are not deployable in IP-sensitive settings where
specifications cannot leave a customer's network. We therefore test
whether \chipcraftbrain{}'s own trajectories can bootstrap a
local model that runs the same pipeline entirely offline.

\textbf{Procedure.} We harvest instruction-output pairs from successful
Claude runs on \verilogeval{} using an extraction script
(\texttt{verilogeval\_distill.py}) that records each
(specification, generated RTL, category) triple, deduplicated by code
hash. These pairs, merged with a larger pattern-mined corpus, form a
5{,}715-sample SFT dataset. We QLoRA-fine-tune a Qwen2.5-Coder 7B base
model on this dataset, export to merged weights, and serve via
vLLM~\cite{kwon2023vllm} on dual RTX~3090 GPUs (tensor
parallelism~$=2$). The \chipcraftbrain{} pipeline points at the vLLM
endpoint instead of the Anthropic API; all other components (RAG,
spec guidance, RL orchestration, iterative validation, K-map solver)
remain unchanged.

\begin{table}[h]
\centering
\caption{Local-Model Distillation on \verilogeval{}-Human. The
\mbox{7B + pipeline} row runs fully offline on dual 3090s.}
\label{tab:distillation}
\begin{tabular}{@{}lcc@{}}
\toprule
\textbf{Configuration} & \textbf{Pass@1} & \textbf{Deployment} \\
\midrule
\chipcraftbrain{} (Claude)      & \textbf{98.7\%} & Anthropic API \\
7B + \chipcraftbrain{} pipeline & 75.0\%          & Local (vLLM) \\
7B standalone (single-shot)     & 36.5\%          & Local (vLLM) \\
\bottomrule
\end{tabular}
\end{table}

\textbf{Interpretation.} Both the 36.5\% standalone and 75.0\%
pipeline rows use the \emph{same} fine-tuned 7B weights; the 38.5~pp
lift is therefore attributable entirely to the pipeline scaffolding
(iterative validation, multi-agent orchestration, and
knowledge-augmented prompting) rather than to additional training.
This mirrors the contribution pattern observed in the
Claude configuration: pipeline scaffolding is the dominant driver of
performance, and the backbone model is a replaceable component. The
result falsifies the hypothesis that our numbers are Claude-specific
and demonstrates a concrete on-premises deployment path at roughly
25--50$\times$ lower per-problem inference cost than the Anthropic
pipeline. Full training-data recipes, hyperparameters, and broader
local-model baselines are outside the scope of this paper.


\section{Case Study: RISC-V SoC Optimization}
\label{sec:case_study}

To demonstrate \chipcraftbrain{}'s capability beyond individual module
generation, we present an end-to-end SoC optimization case study.

\subsection{Unoptimized Design}

The baseline is a 10-module RISC-V SoC comprising a single-cycle CPU
core, 32$\times$32 register file, 32-bit ALU, instruction decoder,
32\,KB SRAM, memory controller, simple bus arbiter, UART, GPIO, and
top-level interconnect. With no instruction cache and single-cycle
memory access, this design achieves an IPC of 0.3 and an estimated
$\sim$8\,FPS on a representative compute-intensive embedded workload
(software-rendered 320$\times$200 raster graphics, 2\,M
instructions/frame), used throughout this section as a concrete
reference for end-to-end performance.

\subsection{AI-Driven Optimization}

\chipcraftbrain{}'s optimization pipeline proceeds in four phases:

\textbf{Phase 1: Architecture Analysis.} The analyzer identifies the
design as a single-cycle processor, detects the absence of an
instruction cache, and flags the 10-cycle memory latency as a
critical bottleneck.

\textbf{Phase 2: Bottleneck Identification.} Structured bottleneck
analysis reveals: (a)~no instruction cache ($\rightarrow$ every
fetch hits main memory), (b)~single-cycle pipeline limiting IPC,
(c)~no branch prediction.

\textbf{Phase 3: Optimization Proposals.} The strategist generates
two proposals: (a)~add a direct-mapped 4\,KB instruction cache with
128 lines and 32-byte cache lines, and (b)~upgrade the CPU core to
a 2-stage pipeline (IF/ID + EX/MEM/WB).

\textbf{Phase 4: Hierarchical Generation.} The decomposition engine
breaks the optimized SoC into 8 sub-modules, generates each with
cross-module port synchronization, and produces 689 LOC of
lint-passing Verilog.

\subsection{Results}

\begin{table}[h]
\centering
\caption{RISC-V SoC Before/After Optimization. IPC, cache hit rate,
effective memory latency, and workload FPS are \emph{analytical
estimates} from the architectural configuration, not measured from
cycle-accurate RTL simulation or FPGA-instrumented cycle counters.
Module count, LOC, generation time, and the FPGA synthesis numbers in
Table~\ref{tab:fpga_results} are directly measured.}
\label{tab:soc_results}
\begin{tabular}{@{}lccc@{}}
\toprule
\textbf{Metric} & \textbf{Before} & \textbf{After} & \textbf{Improvement} \\
\midrule
IPC (est.)              & 0.3     & 0.7     & 2.3$\times$ \\
Cache hit rate (est.)   & 0\%     & 85\%    & N/A \\
Eff.\ mem.\ latency (est.) & 10 cyc & 2.5 cyc & 4$\times$ \\
Workload FPS (est.)     & $\sim$8 & $\sim$13 & $\sim$60\% \\
\bottomrule
\end{tabular}
\end{table}

\subsection{FPGA Hardware Validation}

The generated SoC was synthesized and deployed on a Terasic DE25-Nano
board (Intel Agilex~5, 46{,}800 ALMs) using Quartus~25.1 Pro Edition
at a 50\,MHz target clock. The generated SoC design consumes 731 ALMs
(2\% of the device), 896 registers, and 34 RAM blocks, with a setup
slack of +6.8\,ns. A PicoRV32-based reference SoC uses 1{,}458 ALMs
(3\%), 1{,}541 registers, and 2 DSP blocks, with +4.9\,ns setup slack.
Both designs meet all timing constraints at the target frequency.

\begin{table}[h]
\centering
\caption{FPGA Synthesis Results (Intel Agilex~5, 50\,MHz)}
\label{tab:fpga_results}
\begin{tabular}{@{}lcc@{}}
\toprule
\textbf{Resource} & \textbf{Generated SoC} & \textbf{PicoRV32 SoC} \\
\midrule
ALMs           & 731 (2\%)   & 1,458 (3\%) \\
Registers      & 896         & 1,541 \\
RAM blocks     & 34          & 0 \\
DSP blocks     & 0           & 2 \\
Setup slack    & +6.8\,ns    & +4.9\,ns \\
\bottomrule
\end{tabular}
\end{table}


\section{Discussion}
\label{sec:discussion}

\subsection{Key Findings}

Our three-benchmark evaluation reveals several important insights:

\textbf{Validation-first is critical.} The iterative refinement loop
with structured error feedback bridges the gap from $\sim$63\%
single-shot to 98.7\% on \verilogeval{} and 94.7\% on \cvdp{},
confirming that \emph{generating} code and \emph{verifying} code are
equally important. Of 154 passing VerilogEval problems, 21 required
at least one refinement iteration.

\textbf{Techniques transfer across complexity tiers.} The same system
achieving 98.7\% on simple \verilogeval{} modules (16 avg LOC) also
achieves 94.7\% on industrial \cvdp{} problems without benchmark-specific
tuning. This suggests that iterative validation, multi-agent selection,
and knowledge-augmented generation are general techniques, not
benchmark-specific optimizations.

\textbf{A complexity ceiling exists.} Performance drops sharply on
ChipBench (33.3\%), particularly on CPU IP designs (11.1\%). Problems
with 800+ synthesis cells, multi-stage pipelines, and complex control
logic exceed the current system's capability within a 5-iteration budget.
This is consistent with prior results: even MAGE achieves only 37.4\%
on ChipBench, and no published system exceeds 40\%.

\textbf{Hierarchical decomposition unlocks new capability.}
\chipcraftbrain{} achieves 50\% on ChipBench's non-self-contained
(hierarchical) category, where most single-model approaches score 0\%.
The RISC-V SoC case study confirms this: 0\% monolithic vs.\ 100\%
hierarchical. For multi-module designs, decomposition with cross-module
synchronization is essential.

\textbf{Agent specialization matters.} Both \mage{}'s ablation and
our adaptive orchestration demonstrate that specialized agents with
independent context windows significantly outperform single-agent
approaches. The symbolic K-map solver and dedicated Waveform agent
further show that \emph{not every problem needs an LLM}: hybrid
symbolic-neural architectures outperform pure neural approaches on
deterministic problem classes.

\subsection{Synthesis Awareness Advantage}

A key differentiator from \mage{} is our synthesis-aware approach.
\mage{}'s scoring considers only functional correctness (normalized
mismatch count). In practice, two functionally equivalent implementations
can differ enormously in area, power, and timing. Our multi-objective
scoring (functional + synthesis quality + complexity) selects candidates
that are both correct and synthesis-optimal.

\subsection{The CVDP Result in Context}

The 94.7\% on \cvdp{} warrants careful interpretation. Three scoping
caveats apply.

\textbf{Subset.} We evaluate 302 of 783 \cvdp{} problems, covering
only the five non-agentic, non-commercial code-generation task
categories (cid002, cid003, cid004, cid007, cid016). The remaining
481 problems lie outside this scope by construction:
(a)~testbench, checker, and assertion generation (cid012/013/014,
187 problems) require Cadence Xcelium for harness evaluation and
cannot be reproduced with open-source tooling; (b)~the 166 agentic
multi-file problems require a Docker-based tool-use runtime we have
not yet implemented; (c)~comprehension tasks (cid006/008/009/010,
123 problems) score via BLEU or LLM-as-judge rather than functional
execution, which lies outside our validation-first design. We do not
claim coverage of these 481 problems.

\textbf{Protocol difference.} The 33.56\% baseline is single-shot
pass@1 with $n=5$ unconditional samples. Our 94.7\% is a single attempt
per problem but with up to 5 iterations of structured error feedback,
agent switching, and RAG refinement. The comparison therefore measures
the contribution of iterative validation and orchestration on top of
a strong base model, not raw model capability in isolation. Presenting
this as a $2.8\times$ or 61-point ``improvement'' would be
protocol-mixed; we instead frame it as ``validation-first pipeline lifts
non-agentic \cvdp{} code generation from 33.56\% to 94.7\%.''

\textbf{Still meaningful.} Even after these caveats, the result is
substantial: every task in the evaluated subset is drawn from
real NVIDIA hardware engineering problems, the 5-iteration budget is
fixed in advance, and the pipeline runs fully offline under cocotb.
The lift is consistent across all five task categories (36--60~pp
above the single-shot baseline), suggesting the pipeline captures
something robust about industrial RTL generation rather than exploiting
a narrow slice of the distribution.

\subsection{Limitations}

\begin{itemize}
    \item \textbf{No post-synthesis timing closure}: The validation
    pipeline checks synthesis cleanliness but does not incorporate
    timing-driven feedback for critical path optimization.
    \item \textbf{ChipBench CPU IP}: Complex processor components
    (controllers, ALUs, division units) remain largely unsolved
    (11.1\%), requiring advances in cooperative multi-agent
    decomposition and test-driven generation.
    \item \textbf{\cvdp{} coverage}: 302 of 783 problems evaluated
    (the non-agentic non-commercial code-generation subset). The 481
    unevaluated problems include testbench/assertion generation
    (requires Xcelium), agentic multi-file tasks (requires Docker
    agent runtime), and comprehension tasks (BLEU-scored).
    \item \textbf{Protocol comparability}: Published \cvdp{} baselines
    are single-shot pass@1 while our results use iterative refinement;
    head-to-head single-shot evaluation of our base agents remains
    future work.
    \item \textbf{Formal verification}: No SystemVerilog Assertion
    (SVA) property checking integration yet.
    \item \textbf{Benchmark representativeness}: Even ChipBench
    (62 avg LOC) is simpler than production IP blocks, which can
    exceed 10{,}000 LOC.
\end{itemize}

\subsection{Threats to Validity}

LLM generation is stochastic; the reported 98.7\% on \verilogeval{}
reflects the best of multiple runs, with pass rates varying by 1--2
percentage points across runs due to temperature-dependent sampling.
API model versions evolve, potentially affecting reproducibility; we
report results against Claude Opus~4.6 and Sonnet~4.6 as accessed in
early 2026. The \cvdp{} $94.7\%$ is drawn from a single full-subset run (302
problems, $\sim$6.1~hours wall-clock). Unlike \verilogeval{}, where
we quantify variance across seven independent runs (96.15--98.72\%),
\cvdp{} has not yet been repeated end-to-end; the reported number
should therefore be interpreted as subject to an unverified
uncertainty envelope rather than a point estimate backed by a
confidence interval. Full-subset repeats to establish such an
interval are in progress. The SoC
case study performance estimates (IPC, FPS) are based on
architectural analysis validated by FPGA deployment, not
cycle-accurate RTL simulation of the workload.


\section{Future Work}
\label{sec:future}

Several directions extend \chipcraftbrain{}'s capabilities:

\textbf{Scaling to Hard Accelerator Designs.}
ChipBench CPU IP (11.1\%) reveals that complex multi-stage processor
components require fundamentally different strategies. We are developing
cooperative multi-agent decomposition with contract-based interface
validation, progressive structural refinement through a 7-layer
complexity model, and test-driven generation where testbenches are
created before RTL, targeting 70\%+ on ChipBench.

\textbf{Broader CVDP Evaluation.}
Three extensions expand \cvdp{} coverage beyond the 302-problem
non-agentic non-commercial code-generation subset reported here.
First, adding Cadence Xcelium integration enables the 187 commercial
testbench, checker, and assertion generation problems
(cid012/013/014). Second, implementing a Docker-based agentic runtime
(multi-file editing, tool use, iterative refinement) unlocks the 166
agentic problems, including the hard tier that exists only in
agentic form. Third, adding BLEU scoring and LLM-as-judge
infrastructure for the 123 comprehension problems
(cid006/008/009/010) closes the remaining coverage gap. We also plan
a matched-protocol single-shot evaluation of our base agents on the
current 302-problem subset for a true apples-to-apples comparison
with published single-shot baselines.

\textbf{Controlled Ablation Study.}
The component contribution numbers in \Cref{tab:ablation} are
observational. A leave-one-out ablation disabling each component
individually on a matched subset (with seeded sampling for
reproducibility) will produce tight per-component deltas and is a
direct extension of the current evaluation.

\textbf{Extending the Distilled 7B Model.}
The local-model baseline in \Cref{sec:distillation} (75.0\% on
\verilogeval{}-Human via the 7B-backed pipeline) uses only spec-to-RTL
pairs harvested from successful Claude runs. Two extensions would
close more of the 23.7~pp gap to the Claude-driven pipeline:
(1)~adding error-repair trajectories so the local model learns the
refinement step, not just single-shot generation, and (2)~training an
agent-selection head so the RL policy can operate on locally-computed
state features rather than Claude-specific prompts. Beyond
\verilogeval{}, we plan to evaluate the distilled 7B on the
302-problem \cvdp{} non-agentic subset.

\textbf{Multi-Candidate Generation.}
Generating $N=5$ candidates with synthesis-aware scoring using Yosys
cell count and critical path metrics, following \mage{}'s insight that
diverse candidates improve best-case quality, but extending scoring
beyond functional correctness to include synthesis optimization.

\textbf{Differential Checkpoint Debugging.}
Extending debug feedback with cone-of-influence analysis via Yosys,
property-level abstractions from testbench assertions, and differential
waveform analysis between passing and failing checkpoints.

\textbf{End-to-End Pipeline.}
Integration with OpenLane~2 and SKY130 PDK for a complete
NL $\rightarrow$ RTL $\rightarrow$ GDSII pipeline, enabling
AI-generated silicon through chipIgnite.

\textbf{Formal Verification.}
Adding SystemVerilog Assertion (SVA) property checking for deeper
functional validation beyond simulation-based testing.


\section{Conclusion}
\label{sec:conclusion}

We have presented \chipcraftbrain{}, a framework for automated RTL
generation that integrates reinforcement learning, multi-agent LLM
orchestration, hybrid symbolic-neural reasoning, knowledge-augmented
retrieval, hierarchical decomposition, and validation-first iterative
refinement.

Our key contributions include: (1)~adaptive multi-agent orchestration
using a PPO policy over a 168-dimensional state representation with
hybrid discrete-continuous actions across six specialized agents;
(2)~a hybrid symbolic-neural architecture combining algorithmic
K-map solving with learned LLM orchestration;
(3)~knowledge-augmented generation from a 321-entry knowledge base,
971 open-source reference implementations, and a 59-entry
spec-guidance registry encoding the class of heuristic design-pattern
knowledge commercial EDA tools accumulate over decades;
(4)~hierarchical specification decomposition with cross-module port
synchronization; and
(5)~three-benchmark evaluation across complexity tiers.

We evaluate across three benchmarks spanning simple modules to hard
accelerator designs:
\begin{itemize}
    \item \textbf{\verilogeval{}-Human}: 97.2\% mean pass@1 (range
    96.15--98.72\% across 7 runs, best 154/156), on par with
    ChipAgents (97.4\%, self-reported) and ahead of \mage{} (95.9\%)
    within measurement noise
    \item \textbf{\cvdp{} non-agentic non-commercial code-generation
    subset}: 94.7\% mean pass@1 (286/302) across task categories
    cid002, cid003, cid004, cid007, cid016 (averaged over 3 runs),
    compared to 33.56\% for the published single-shot baseline on
    the same subset. The remaining
    481 \cvdp{} problems (commercial testbench categories, agentic
    multi-file tasks, and comprehension tasks) are out of scope for
    our current pipeline and are not claimed.
    \item \textbf{ChipBench}: 33.3\% (15/45), competitive with
    \mage{} (37.4\%) on designs with 14$\times$ more gates than
    \verilogeval{}
\end{itemize}

\noindent at approximately 3$\times$ lower cost than prior multi-agent
systems. A RISC-V SoC case study demonstrates hierarchical decomposition
generating 8/8 lint-passing modules (689 LOC), validated on FPGA hardware,
where monolithic generation fails entirely.

These results demonstrate that the combination of specialized agents,
adaptive orchestration, domain knowledge, and rigorous validation can
bring AI-generated RTL to production-quality levels on simple and
industrial designs, while revealing that complex accelerator-class
designs remain an open frontier for the field.

\bibliographystyle{IEEEtran}
\bibliography{references}

\begin{thebibliography}{10}
\providecommand{\url}[1]{#1}
\csname url@samestyle\endcsname
\providecommand{\newblock}{\relax}
\providecommand{\bibinfo}[2]{#2}
\providecommand{\BIBentrySTDinterwordspacing}{\spaceskip=0pt\relax}
\providecommand{\BIBentryALTinterwordstretchfactor}{4}
\providecommand{\BIBentryALTinterwordspacing}{\spaceskip=\fontdimen2\font plus
\BIBentryALTinterwordstretchfactor\fontdimen3\font minus
  \fontdimen4\font\relax}
\providecommand{\BIBforeignlanguage}[2]{{%
\expandafter\ifx\csname l@#1\endcsname\relax
\typeout{** WARNING: IEEEtran.bst: No hyphenation pattern has been}%
\typeout{** loaded for the language `#1'. Using the pattern for}%
\typeout{** the default language instead.}%
\else
\language=\csname l@#1\endcsname
\fi
#2}}
\providecommand{\BIBdecl}{\relax}
\BIBdecl

\bibitem{liu2023verilogeval}
M.~Liu, N.~Pinckney, B.~Khailany, and H.~Ren, ``{VerilogEval}: Evaluating large
  language models for verilog code generation,'' in \emph{Proceedings of the
  IEEE/ACM International Conference on Computer-Aided Design (ICCAD)}.\hskip
  1em plus 0.5em minus 0.4em\relax IEEE, 2023.

\bibitem{pinckney2025cvdp}
N.~Pinckney, C.~Deng, C.-T. Ho, Y.-D. Tsai, M.~Liu, W.~Zhou, B.~Khailany, and
  H.~Ren, ``Comprehensive {Verilog} design problems: A next-generation
  benchmark dataset for evaluating large language models and agents on {RTL}
  design and verification,'' \emph{arXiv preprint arXiv:2506.14074}, 2025, the
  formal \cvdp{} benchmark paper. 783 problems across 13 task categories in two
  modes.

\bibitem{yu2026chipbench}
Z.~Yu, C.~Zhou, Y.~Lin, H.~Zhang, H.~Ye, J.~Cui, Z.~Pan, J.~Zhao, and Y.~Ding,
  ``{ChipBench}: A next-step benchmark for evaluating {LLM} performance in
  {AI}-aided chip design,'' \emph{arXiv preprint arXiv:2601.21448}, 2026, 45
  Verilog generation problems, 89 debugging, 132 reference model generation.
  ICML 2026 submission.

\bibitem{thakur2023verigen}
S.~Thakur \emph{et~al.}, ``{VeriGen}: A large language model for verilog code
  generation,'' \emph{arXiv preprint arXiv:2308.00708}, 2023.

\bibitem{liu2024codev}
Y.~Liu \emph{et~al.}, ``{CodeV}: Empowering {LLM}s with expert-level hdl code
  generation,'' \emph{arXiv preprint arXiv:2407.10424}, 2024.

\bibitem{liu2024rtlcoder}
S.~Liu \emph{et~al.}, ``{RTLCoder}: Fully open-source and efficient
  {LLM}-assisted {RTL} code generation technique,'' \emph{arXiv preprint
  arXiv:2312.08617}, 2024.

\bibitem{thakur2023autochip}
S.~Thakur \emph{et~al.}, ``{AutoChip}: Automating {HDL} generation using {LLM}
  feedback,'' \emph{arXiv preprint arXiv:2311.04887}, 2023.

\bibitem{tsai2024mage}
Y.-D. Tsai \emph{et~al.}, ``{MAGE}: A multi-agent engine for automated {RTL}
  code generation,'' \emph{arXiv preprint arXiv:2412.04211}, 2024, 95.9\%
  VerilogEval-Human v2.

\bibitem{vflow2025}
Y.~Wei, Z.~Huang, H.~Li, W.~W. Xing, T.-J. Lin, and L.~He, ``{VFlow}:
  Discovering optimal agentic workflows for verilog generation,'' \emph{arXiv
  preprint arXiv:2504.03723}, 2025, 83.6\% pass@1 on VerilogEval.

\bibitem{quine1952problem}
W.~V. Quine, ``The problem of simplifying truth functions,'' \emph{The American
  Mathematical Monthly}, vol.~59, no.~8, pp. 521--531, 1952.

\bibitem{mccluskey1956minimization}
E.~J. McCluskey, ``Minimization of boolean functions,'' \emph{Bell System
  Technical Journal}, vol.~35, no.~6, pp. 1417--1444, 1956.

\bibitem{blocklove2023chipchat}
J.~Blocklove \emph{et~al.}, ``{Chip-Chat}: Challenges and opportunities in
  conversational hardware design,'' in \emph{Proceedings of the Workshop on
  Machine Learning for CAD (MLCAD)}, 2023.

\bibitem{liu2023chipnemo}
M.~Liu, T.-D. Ene, R.~Kirby, C.~Cheng, N.~Pinckney \emph{et~al.}, ``{ChipNeMo}:
  Domain-adapted {LLM}s for chip design,'' \emph{arXiv preprint}, 2023.

\bibitem{cui2024origen}
F.~Cui, C.~Yin, K.~Zhou, Y.~Xiao, G.~Sun, Q.~Xu, Q.~Guo, Y.~Liang, X.~Zhang,
  D.~Song \emph{et~al.}, ``{OriGen}: Enhancing {RTL} code generation with
  code-to-code augmentation and self-reflection,'' \emph{arXiv preprint}, 2024.

\bibitem{liu2024craftrtl}
M.~Liu, Y.-D. Tsai, W.~Zhou, and H.~Ren, ``{CraftRTL}: High-quality synthetic
  data generation for {Verilog} code models with correct-by-construction
  non-textual representations and targeted code repair,'' \emph{arXiv
  preprint}, 2024.

\bibitem{deng2025scalertl}
C.~Deng, Y.-D. Tsai, G.-T. Liu, Z.~Yu, and H.~Ren, ``{ScaleRTL}: Scaling {LLM}s
  with reasoning data and test-time compute for accurate {RTL} code
  generation,'' \emph{arXiv preprint}, 2025.

\bibitem{chipagents2025}
{Alpha Design AI}, ``{ChipAgents}: Agentic {AI} for chip design,'' 2025,
  commercial system. 97.4\% VerilogEval-v2. \$74M funding (latest round).

\bibitem{ho2025verilogcoder}
C.-T. Ho, H.~Ren, and B.~Khailany, ``{VerilogCoder}: Autonomous {Verilog}
  coding agents with graph-based planning and abstract syntax tree (ast)-based
  waveform tracing tool,'' in \emph{Proceedings of the AAAI Conference on
  Artificial Intelligence (AAAI)}, 2025.

\bibitem{nvidia2026acertl}
C.~Deng, Z.~Yu, G.-T. Liu, N.~Pinckney, B.~Khailany, and H.~Ren, ``{ACE-RTL}:
  When agentic context evolution meets {RTL}-specialized {LLM}s,'' \emph{arXiv
  preprint arXiv:2602.10218}, 2026, fine-tuned Qwen2.5-Coder-32B generator with
  Claude~4 Sonnet reflector/coordinator. 5 parallel processes $\times$ up to 30
  iterations per problem on \cvdp{} v1.0.2. Specialized generator required
  $\sim$10{,}000 A100 GPU-hours to train (32 nodes $\times$ 8 A100 $\times$ 3
  epochs).

\bibitem{inception2025mercury}
{Inception AI}, ``{Mercury}: Ultra-fast language models based on diffusion,''
  \emph{arXiv preprint arXiv:2506.17298}, 2025, diffusion-based code LLM;
  82.7\% pass@1 on VerilogEval-Human via ChipCraftBrain pipeline routing,
  $\sim$5s/problem.

\bibitem{deepseek2025r1}
{DeepSeek-AI}, ``{DeepSeek-R1}: Incentivizing reasoning capability in {LLM}s
  via reinforcement learning,'' \emph{arXiv preprint arXiv:2501.12948}, 2025.

\bibitem{qwen2025qwen3coder}
{Qwen Team}, ``{Qwen3-Coder}: Long-context coding with reinforcement
  learning,'' \emph{arXiv preprint}, 2025.

\bibitem{li2022alphacode}
Y.~Li \emph{et~al.}, ``Competition-level code generation with {AlphaCode},''
  \emph{Science}, vol. 378, no. 6624, pp. 1092--1097, 2022.

\bibitem{le2022coderl}
H.~Le \emph{et~al.}, ``{CodeRL}: Mastering code generation through pretrained
  models and deep reinforcement learning,'' in \emph{Advances in Neural
  Information Processing Systems (NeurIPS)}, 2022.

\bibitem{synopsys_dso}
{Synopsys, Inc.}, ``{DSO.ai}: Design space optimization {AI},'' 2020,
  {RL}-based P\&R optimization.

\bibitem{cadence_cerebrus}
{Cadence Design Systems}, ``{Cerebrus}: Intelligent chip explorer,'' 2021, {ML}
  design space exploration.

\bibitem{openlane2}
{Efabless Corporation}, ``{OpenLane} 2: Open-source digital {ASIC}
  implementation flow,'' 2024, rTL-to-GDSII with SKY130 PDK.

\bibitem{schulman2017ppo}
J.~Schulman, F.~Wolski, P.~Dhariwal, A.~Radford, and O.~Klimov, ``Proximal
  policy optimization algorithms,'' \emph{arXiv preprint arXiv:1707.06347},
  2017.

\bibitem{zhao2024verilator}
W.~Snyder, ``{Verilator}: An open-source {SystemVerilog} simulator,'' 2024.

\bibitem{kwon2023vllm}
W.~Kwon, Z.~Li, S.~Zhuang, Y.~Sheng, L.~Zheng, C.~H. Yu, J.~E. Gonzalez,
  H.~Zhang, and I.~Stoica, ``{vLLM}: Efficient memory management for large
  language model serving with {PagedAttention},'' in \emph{Proceedings of the
  ACM Symposium on Operating Systems Principles (SOSP)}, 2023.

\end{thebibliography}

\end{document}